\documentclass[aps,prl,twocolumn,groupedaddress,superscriptaddress,longbibliography]{revtex4}

\usepackage{amsmath}
\usepackage{color}
\usepackage{amsfonts}
\usepackage{epsf}
\usepackage{graphicx}
\usepackage{mathrsfs}
\usepackage[caption=false]{subfig}

\definecolor{red}{rgb}{1,0,0}
\definecolor{blue}{rgb}{0,0,1}

\usepackage{mathrsfs}
\begin{document}

\title{Transcriptional bursts in a non-equilibrium model for gene regulation by supercoiling}

\author{M. Ancona}
\affiliation{SUPA, School of Physics and Astronomy, University of Edinburgh, Peter Guthrie Tait Road, Edinburgh, EH9 3FD, UK}
\author{A. Bentivoglio}
\affiliation{SUPA, School of Physics and Astronomy, University of Edinburgh, Peter Guthrie Tait Road, Edinburgh, EH9 3FD, UK} 
\author{C.~A. Brackley}
\affiliation{SUPA, School of Physics and Astronomy, University of Edinburgh, Peter Guthrie Tait Road, Edinburgh, EH9 3FD, UK}
\author{G. Gonnella}
\affiliation{Dipartimento di Fisica, Universit\`a di Bari and INFN, Sezione di Bari, 70126 Bari, Italy}
\author{D. Marenduzzo}
\affiliation{SUPA, School of Physics and Astronomy, University of Edinburgh, Peter Guthrie Tait Road, Edinburgh, EH9 3FD, UK}

\newcommand{\CBcomment}[1]{\textcolor{blue}{Chris: \textit{#1}}}
\newcommand{\Marco}[2]{\textcolor{blue}{Marco: \textit{#1}}}
\newcommand{\SM}[0]{Ref.~\cite{SM}}
\newcommand{\red}[1]{{\color{red} #1}}
\newcommand{\green}[1]{{\color{green} #1}}

\begin{abstract}
We analyse transcriptional bursting within a stochastic non-equilibrium model which accounts for the coupling between the dynamics of DNA supercoiling and gene transcription. We find a clear signature of bursty transcription when there is a separation between the timescales of transcription initiation and supercoiling dissipation -- the latter may either be diffusive or mediated by topological enzymes, such as type I or type II topoisomerases. In multigenic DNA domains we observe either bursty transcription, or transcription waves; the type of behaviour can be selected for by controlling gene activity and orientation. In the bursty phase, the statistics of supercoiling fluctuations at the promoter are markedly non-Gaussian. 
\end{abstract}

\maketitle

\section*{Introduction} 

Transcription, the mechanism through which DNA is read by a polymerase to create a messenger RNA, is a crucial process in living cells, and its regulation is an important determinant of cell function~\cite{Alberts}. The dynamics of transcription are inherently stochastic, since its initiation requires an RNA polymerase (RNAP) and associated co-factors to bind at the promoter of a gene~\cite{Raj}. Not only can the copy numbers of RNAPs and transcription factors be low, but the mechanism through which they find their specific binding sites (a combination of 3D diffusion through the nucleoplasm or cytoplasm, and 1D diffusion along the genome~\cite{Berg,Marko}) leads to a broad distribution of search times. As a consequence, transcription initiation is a rare event with rates of the order of inverse hours in mammals~\cite{Schwanhausser}, while in bacteria tens of minutes may elapse between successive initiation events at the same promoter~\cite{Golding}.

An intriguing feature of transcription as a stochastic process is that it is often ``bursty''. This means that when recording transcription events for a given gene in a given cell, one observes clusters of closely spaced events separated by longer dormant periods in which the gene is silent, and the distribution of interval times is non-Poissonian. Transcription bursts are common to both bacterial and eukaryotic cells; this phenomenon is thought to provide a potential basis for the variability in behaviour which is observed in genetically identical cells within the same environment~\cite{Raj}. Bursting may therefore play a key role in the pathway through which a cell can spontaneously break symmetry, for instance to choose its fate early in development in higher eukaryotes~\cite{Losick}.

From a general point of view, a dynamical system yields bursty behaviour when there is intermittent switching between states with high and low activity~\cite{Gardiner,Sneppen}. But what is the biophysical mechanism underlying such intermittency? Several possibilities have been proposed in the literature, and they may differ in different organisms. In mammals, proposed mechanisms for bursting are typically gene-specific~\cite{Suter}; cycles of higher transcription activity may be due to chromatin remodelling~\cite{Harper}, or to the action of cis-regulatory DNA elements~\cite{Suter}. In both cases, refractory periods during which genes are silent last for hours. In bacteria, refractory timescales only span tens of minutes, and other scenarios may be more relevant. For instance, pausing of an RNAP along a gene may lead to the formation of a queue of multiple RNAPs behind it, producing bursts of transcripts~\cite{Corrigan,Levine2018}. This mechanism requires that the gene in question is highly expressed, so that multiple polymerases can be recruited to its promoter within minutes. Recently, a set of experiments~\cite{Chong} has demonstrated that transcriptional bursts in DNA plasmids {\it in vitro} are associated with DNA supercoiling~\cite{Maxwell}, which is the extent of over or under-winding of the two strands in a DNA double helix. Specifically, it was shown that the build-up of positive supercoiling (DNA over-twisting) stalls transcription, and the addition of gyrase (which relaxes this supercoiling) leads to transcriptional bursts~\cite{Chong}. The conclusions of this work are consistent with the hypothesis in the earlier rate-based theoretical model in Ref.~\cite{Sneppen}, which suggested that the timescales observed in bacterial bursts are compatible with supercoiling-dependent initiation.

Here we consider a stochastic model which couples the dynamics of supercoiling and transcription in DNA (first introduced in Ref.~\cite{Brackley}), and ask whether, and under which conditions, supercoiling may lead to transcriptional bursts. Our main result is that supercoiling can induce bursts in a wide range of parameter space. Perhaps surprisingly, bursts occur only when the overall transcriptional rate is low, and are absent when it is high. When gene density is low (e.g., if we model a single gene), significant bursting is primarily found in the presence of topological enzymes which relax supercoiling at a fixed rate. For higher gene density, highly significant bursts can also arise in the absence of topoisomerases, through the action of a self-organised non-equilibrium regulatory network mediated by supercoiling. Intriguingly, this same pathway can also generate transcription waves and upregulate divergent transcription, both of which eventually disrupt the bursty behaviour. A final key finding is that bursting leaves a detectable signature in the distribution of supercoiling at the promoter -- it leads to a non-normal distribution, and the appearance of a singularity or small peak in the tails of the distribution. Observation of such signatures may be an aim of future experiments -- along the lines of existing ones, performed either {\it in vitro}~\cite{Chong} or {\it in vivo}~\cite{Naughton,Kouzine}. 

\section*{Materials and methods}

\subsection*{Model}

The key dynamical variable in our model is a one-dimensional scalar field $\sigma(x,t)$, which denotes the local supercoiling density at a point $x$ on the DNA.  This is the local analog of the global supercoiling density, defined as $({Lk} - \mbox{Lk}_0 )/\mbox{Lk}_0$, with $\mbox{Lk}$ and $\mbox{Lk}_0$ respectively the linking number -- which can be decomposed into twist and writhe~\cite{Maxwell} -- of a DNA molecule and that of a torsionally relaxed B-DNA, that is $1$ for every $10.5$ base pairs (bp)~\footnote{As both $\mbox{Lk}$ and $\mbox{Lk}_0$ can be defined for a segment of the DNA molecule, an analogous formula can be used to define $\sigma$ locally -- doing this is analogous to approximating writhe with a local field~\cite{Kamien}.} The dynamics are then described in continuous space by
\begin{equation}
\frac{\partial \sigma(x,t)}{\partial t} = \frac{\partial}{\partial x}\left[D \frac{\partial}{\partial x} \sigma(x,t) - J_{\rm tr}(x,t)\right] - k_{\rm topo}\sigma(x,t),
\label{seq:eqModel}
\end{equation}
where the three terms on the right-hand side represent diffusion of supercoiling, supercoiling flux generated by transcription, and supercoiling dissipation due to topological enzymes. Below we will discuss each term in turn. Though we have written Eq.~(\ref{seq:eqModel}) as continuous in space, we solve it on a lattice of length $L=15$~kbp with spacing $\Delta x=15$~bp (which is approximately the footprint size of an RNAP).

If we consider a closed DNA loop, then in the absence of topological enzymes the total level of supercoiling is conserved (it is a topological invariant of the system). Therefore we require that supercoiling obeys ``model B'' (conserved) dynamics~\cite{Chaikin1995}. Further, the free energy density of supercoiling (twist and writhe) $f$ is, to a good approximation, quadratic in the supercoiling density~\cite{Maxwell,Brackley}; so the chemical potential ($\partial f/\partial \sigma$) is linear in $\sigma$. Since the flux in model B dynamics is proportional to the gradient of the chemical potential, this results in a diffusion equation (when the mobility is constant)~\cite{Brackley}, giving the first term on the right of Eq.~(\ref{seq:eqModel}). Single molecule measurements of the dynamics of plectonemic supercoils \textit{in vitro}~\cite{vonLoenhout2012} and studies of transcription {\it in vivo}~\cite{Moulin} are also consistent with a diffusive dynamics for supercoiling.

The second term in Eq.~(\ref{seq:eqModel}) represents supercoiling fluxes due to transcription. Through this term we couple the supercoiling dynamics to stochastic transcriptional kinetics, where each of $N$ RNAPs can bind at the promoters of $n$ genes (each of size $ \lambda = 66\Delta x \sim 1000 $bp) located at lattice positions $y_j$, $j=1,\ldots,n$. Transcription initiates stochastically when an inactive RNAP binds at gene $j$ with rate $k_{{\rm in},j}$. The RNAP then elongates with velocity $v$ (positive or negative depending on the direction of transcription) such that at a time $t_i$ after initiating it is located at position $x_i(t_i)=y_j+vt_i$ (where the index $i$ labels the RNAP). Transcription terminates (and the promoter becomes again available for initiation) once the RNAP reaches the end of a gene. The total flux is then given by 
\begin{equation}
J_{\rm tr}(x,t) = \sum_{i=1}^N J_i(t_i) \delta(x - x_i(t_i)) \eta_i(t),
\label{seq:flux}
\end{equation}
where the sum is over all RNAPs, $J_i(t_i)$ is the flux generated by RNAP $i$, and the function $ \eta_i(t) $ represents its state, taking a value of $0$ if it is unbound, and $1$ if it is actively transcribing. The initiation dynamics is coupled back to the supercoiling by making $k_{{\rm in},j}$ a function of the supercoiling at the promoter,
\begin{equation}
k_{{\rm in},j}(t)=k_0{\rm max}\left\{1-\alpha \sigma(x_j,t),0\right\},
\label{seq:rate}
\end{equation}
where $\alpha$ is a coupling parameter (it represents the sensitivity of RNAP-DNA binding to the level of supercoiling).

According to the {\it twin supercoiling domains} model~\cite{Liu}, a moving RNAP generates a supercoiling flux if its rotation is hindered, because as the enzyme progresses the DNA has to locally unwind. [The rotational drag on the RNAP \textit{in vivo} is likely to be large in view of its size and of its interactions with other macromolecules.] As a result, positive supercoiling is generated in front of the RNAP, and negative supercoiling behind. More specifically, as the RNAP moves, all the DNA twist in front of it (minus any rotation of the RNAP if present) is pushed forward. This reasoning suggests that the flux generated will depend on the level of twist ahead of the RNAP, which might in principle vary~\cite{Levine2018}. To obtain a tractable model we assume though this flux to be constant. In practice, this approximation is likely good as DNA can only support small levels of twist before writhing~\cite{Marko2007}.

A final complication is that diffusion of supercoils through the RNAP also requires its rotation -- hence, we expect this effect to be small. We could prohibit flux through the RNAPs by introducing no-flux boundary conditions at the points $x_i$, but an alternative which yields a more tractable model is to instead ramp up the flux as transcription progresses; to do this we set
\begin{equation}
J_i(t_i) = J_0\left(1+\frac{|v|t_i}{\Delta x}\right).
\label{seq:modflux}
\end{equation} 
The sign of $ J_0 $ depends on the direction of transcription. Any residual net diffusive leak only plays a minor role, as it is small compared to $J_0$. [Additionally, a small leak of supercoiling may not be unrealistic even for a polymerase acting as a topological barrier for twist, if the region of DNA containing it writhes in 3D.] In what follows, we define $\bar{J}=J_0\left[1+\lambda/(2\Delta x\right)]$ -- this is a useful quantity as it is the average value of the supercoiling flux generated during a transcription event.

The third term on the right of Eq.~(\ref{seq:eqModel}) represent loss of supercoiling due to the action of topoisomerases (such as, e.g., topoI, topoII, gyrase). This is introduced in the model in a minimal way, as a first-order reaction where both positive and negative supercoiling relax at the same rate $ k_{\rm topo} $. In general this term does not conserve the total supercoiling; however here we start with a uniform initial condition $\sigma(x,t = 0) \equiv \sigma_0$ where $\sigma_0=0$, so in this case the total supercoiling is conserved. 

This model was first described in Ref.~\cite{Brackley}. In that work it was found that by increasing the ratio $\bar{J}/D$ there is a crossover from a relaxed regime where transcription is virtually Poissonian, to a supercoiling-regulated regime where transcription of neighbouring genes is highly correlated. In the present work we instead ask whether, and under what conditions, the coupling between supercoiling and transcription can lead to bursty dynamics. 

\subsection*{Key model quantities and parameter values}

Key quantities which control the model behaviour are the ratios $\bar{J}/D$ and $k_{\rm topo}/k_0$. In the results section below we will explore the ability of the model to exhibit bursty behaviour at different points within the $\bar{J}/D$--$k_{\rm topo}/k_o$ parameter space, for different gene arrangement cases. Though the values which these quantities have \textit{in vivo} have not yet been well characterised experimentally, here we discuss what ranges of values might be relevant based on available evidence. For our simulations we vary several parameters in order to get an understanding of how such a system might behave under different conditions.

The diffusion constant for supercoiling is difficult to measure -- not least because one would expect very different values for twist and writhe. Intuitively one would expect twist to diffuse very quickly~\cite{Levine2018}, whereas writhe diffusion will be much slower, since it requires more global DNA rearrangements. Also, it has been shown that DNA is unable to support much deviation of twist from its relaxed state: the theory in Ref.~\cite{Marko2007} (and Refs. therein) indicate that it will writhe if the supercoiling density exceeds $0.01$. This suggests that the slower diffusion of writhe will dominate the dynamics. Single molecule experiments presented in Ref.~\cite{vonLoenhout2012}, which measured the motion of plectonemes in a stretched DNA molecule, indeed obtained a relatively small diffusion coefficient, with a value significantly less than $1$ kbp$^2$/s (see Fig.~3F in~\cite{vonLoenhout2012}). Specifically, when a DNA molecule is subjected to tensions of less than $1$--$2$~pN, plectoneme diffusivity is at most $\sim 0.1$ kbp$^2$/s. Taking this value, and a typical size for a bacterial gene of $\lambda\sim 1$~kbp, the time it takes for supercoiling to diffuse away from the promoter after transcription terminates is $\sim \lambda^2/(2D)\sim 5$~s. \textit{In vivo}, macromolecular crowding is likely to further slow down writhe/supercoiling diffusion -- so for our simulations we consider values for  $D$ which are between $\sim 4$ and $\sim 40$ times smaller than the value quoted above. Specifically, we consider $D\sim 2.25\times 10^{-2}$ kbp$^2$/s in Figure~\ref{fig:figure1}, and $D\sim 2.25\times 10^{-3}$~kbp$^2$/s in other figures. Both values allow the supercoiling generated during transcription of a gene to dissipate in at most minutes after transcription termination (the smaller value was used for the more systematic analyses as it enables more efficient simulations -- as it is possible to use a larger value for the time step).

The typical RNAP velocity in bacteria is $\sim 100$ bp/s~\cite{Alberts}, so that the time taken to transcribe a $\lambda=1$~kbp gene is $\tau \sim \lambda/v\sim 10$~s. 
Then, through dimensional analysis, we expect an order-of-magnitude estimate for the $\bar{J}$ to be $\sim \lambda^2/\tau$ or $v\lambda\sim 0.1$~kbp$^2$/s -- notably, this is the same order of magnitude as $D$. Thus in our simulations we explore a range of values for $\bar{J}/D$ which is typically between $0.34$ and $3.4$.

Turning now to the dynamics of transcription initiation, measured RNAP initiation rate can vary widely, and typical values are in the range $1$ s$^{-1}$ to $1$ hr$^{-1}$ are observed (see~\cite{Brackley,Liang99,yeasttranscription,humantranscription}). Likewise the number of RNAPs has a large variability. For example in the bacteria {\it E. coli} there are an estimated $1000$--$10000$ RNAP per cell~\cite{Bremer1996}, and about $5000$ genes. To reflect this ratio, we take one RNAP per gene in our simulations, unless otherwise stated (see Supporting Material, Fig.~S4). The rate of topoisomerase action {\it in vivo} is equally difficult to estimate. Ref.~\cite{Ishihama2008} counts $\sim 500$ topoisomerase I per cell in {\it E. coli}. Assuming that about half enzyme are bound, and that there are two genomes per cell on average, we arrive at $\sim 100$ topoisomerase proteins are bound per genome, or $0.02$ per gene. Assuming additionally that each enzyme can on average relax $1$--$10$ supercoil per second~\cite{Terekhova2012}, and that the baseline bacterial supercoiling is equal to $-0.05$, we get $k_{\rm topo}\sim 0.005$--$0.05$ s$^{-1}$. We note this rough estimate is within the physiological value of the baseline transcriptional rate (in our simulations $k_{0}$). Due to this, and since the values of  $k_{\rm topo}$ and $k_{0}$ are not known accurately, in our simulations we have systematically varied their ratio $k_{\rm topo}/k_{0}$, between $0$ and $1.4$. [Specifically we set $k_0=0.001$ s$^{-1}$, and varied $k_{\rm topo}$.] In this way we can examine all possible scenarios concerning the balance between initiation and topoisomerase relaxation rates.

To determine whether supercoiling can affect RNAP initiation at a promoter, we need to consider the time it takes for supercoiling generated by a previous transcription event to diffuse away, and the typical initiation rate ($k_{{\rm in},j}$ in our model). In Ref.~\cite{Brackley} a mean field model was used to estimate the extent of residual supercoiling at the promoter, and this was given by 
\begin{equation}
|\sigma_p| \simeq \frac{\bar{J}}{2D} k_0\tau.
\end{equation}
If this quantity is larger than $\alpha^{-1}$ (see Eq.~\eqref{seq:rate} for the definition of $\alpha$), then supercoiling can indeed increase the rate at which RNAP binds the promoter, leading to a positive feedback. Experiments in bacterial genes suggest that a supercoiling density of $\sigma_p\leq-0.01$ is sufficient to enhance RNAP binding (see~\cite{Brackley,Weintraub1986,Rhee1999}), so in our simulations we have set $\alpha=100$. These considerations explain why $\bar{J}/D$ and $k_{in} \tau$ are key dimensionless quantities in our model.

In what follows we give parameter values either in physical units, or in terms of dimensionless ratios. When required, the physical values of all parameters can be reconstructed by referring back to this section. As noted above, to solve Eq.~(\ref{seq:eqModel}) numerically we discretise space into a lattice with 15~bp spacing; we discretise time into steps of between $0.1$--$1$~s, chosen to balance efficiency and numerical stability (which depends on the other parameters). 

\subsection*{The sequence-size function}

\begin{figure}[t]
\centering
{\includegraphics[width=.5\textwidth]{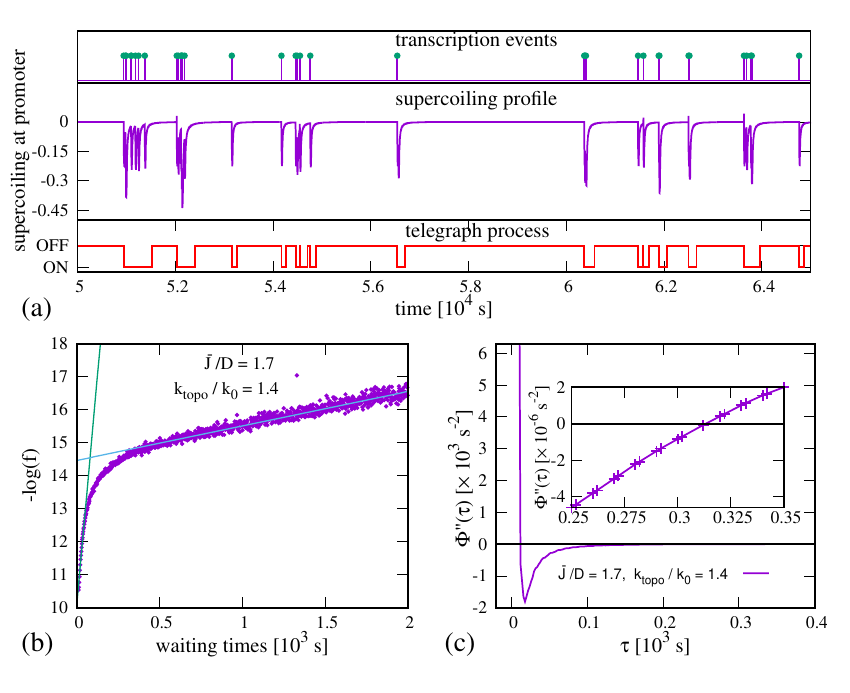}} \\
\caption{\textbf{Transcriptional bursts and the sequence-size function}. \textbf{(a)} Time series for a simulation with $ \bar{J}/D = 1.7 $, $ k_{\rm topo}/k_0 = 1.4 $. Transcription events (top) are often grouped in bursts. Transcription initiation depends on the level of supercoiling at the gene promoter (middle). The state of the system (bottom) is defined as OFF if $ \sigma_p \geq (1-(k_0 \tau_2)^{-1})/\alpha $, and ON otherwise. \textbf{(b)} Negative logarithm of the \textit{pdf} of waiting times. The existence of two linear regions characterises the dynamics as bursty.
\textbf{(c)} Second derivative of the \textit{ssf}: the existence of roots at $ \tau_1 $ and $ \tau_2 $ demonstrates the presence of two timescales. Inset: zoom of the intersection of this second derivative with the x-axis close to $ \tau_2 $.}
\label{fig:figure1}
\end{figure}

A classic model to describe transcriptional bursting is the \textit{interrupted Poisson process} (IPP)~\cite{Dobrzynsky} (see Supporting Material, Section~II), which describes transitions between an active (ON) and an inactive (OFF) state with Poissonian rates $k_{\rm ON} $ and $ k_{\rm OFF} $, as in a random telegraph process~\cite{Gardiner}, together with transcription at a constant rate $k_i$ whilst the system is in the ON state. The process can be characterised by the probability distribution function (\textit{pdf}) $f(t)$ of waiting times -- the time intervals between two consecutive transcriptional events --  which is given by a double-exponential~\cite{Dobrzynsky}, where the two characteristic times are related to the interval between transcriptions in a single burst, and the interval between two consecutive bursts (see Supporting Material, Eqs.~(S10)--(S13)).

To determine whether our system is bursty for a given set of parameters ($ \bar{J}/D $ and $k_{\rm topo}/k_0$) we measure the distribution $f(t)$. We then analyse the so-called \textit{sequence-size function} (\textit{ssf})~\cite{Dobrzynsky,Kumar}, which is defined in terms of $f(t)$ as 
\begin{equation}
\Phi(\tau) = \frac{1}{1-\int_0^{\tau} f(t)dt}.
\label{eq:doubleexp}
\end{equation}
This is the inverse of the probability of observing a waiting time larger than $\tau$, or, equivalently, the proportion of transcriptional event intervals which are longer than $\tau$. 
If the dynamics is bursty, we expect two well-separated timescales for the decay of $f(t)$; correspondingly  $\Phi$ will display a plateau and two inflection points. These points, $\tau_1$ and $\tau_2>\tau_1$, can be found as the zeros in the second derivative of $\Phi$; these values also approximate the two timescales for $f(t)$ (this is not a strict equality, but rather an order-of-magnitude estimate). The value of $\Phi$ in the middle of the plateau, ($ \Phi(\tau_x) = (\Phi(\tau_1)+\Phi(\tau_2))/2 $), then yields the average number of transcriptional events in a burst, or burst size, $\beta$~\cite{Dobrzynsky}. 
If the dynamics is not bursty, $\Phi$ will have no more than one inflection point. An analysis of $f(t)$ and $\Phi(t)$ shows that these criteria work qualitatively well for our model. For sufficiently large values of $k_{\rm topo}$, we find that the dynamics is indeed bursty (Fig.~1): $f(t)$ has two characteristic timescales (Fig.~\ref{fig:figure1}b), and $\Phi(\tau)$ has two well defined inflection points (Fig.~\ref{fig:figure1}c). 
For small $k_{\rm topo}$ and high values of the flux, there are no well defined timescales in $f(t)$ or inflection points in $\Phi(\tau)$; inspection of the transcriptional dynamics show this is not bursty (see Supporting Material, Fig.~S3c).

To quantify the ``burstiness'' of a transcriptional time series, we define the following parameter,  
\begin{equation}
\xi = \frac{\Phi'(\tau_1) - \Phi'(\tau_2)}{\Phi'(\tau_1)},
\label{eq:significance}
\end{equation}
which measures the area under $\Phi''(\tau)$ between the two inflection points (when they exist), normalised by $\Phi'(\tau_1)$ so that the result remains between $0$ and $1$ (prime and double prime denote first and second derivatives respectively). 
$\xi$ is zero when the dynamics are not bursty, and increases as the separation between the two characteristic timescales $\tau_1$ and $\tau_2$ becomes clearer: we refer to this parameter as the \textit{burst significance}.

\section*{Results}

\subsection*{A single gene}

\begin{figure}[t]
\centering
{\includegraphics[width=.5\textwidth]{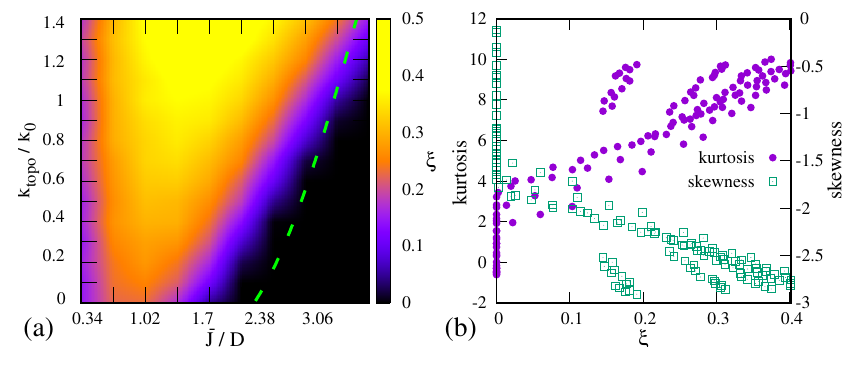}} \\
\caption{\textbf{Burstiness for a single gene}. \textbf{(a)} Phase diagram showing burst significance as a function of model parameters. A \textit{non-bursty} regime is indicated by $\xi\sim 0$ (black) while $\xi>0$ indicates a \textit{bursty} regime (yellow-red). The dashed green line is the boundary predicted via mean field (see Supporting Material, Section~I). \textbf{(b)} Non-Gaussian parameters of the distribution of the supercoiling at the promoter $\sigma_p$ as function of $\xi$, for different values of $ \bar{J}/D $ and $ k_{\rm topo}/k_0 $. The kurtosis (skewness) is correlated (anticorrelated) with $ \xi $.} 
\label{fig:figure2}
\end{figure}

We first consider the case of a single gene. By computing $\xi$ from simulations with different values of $ \bar{J}/D $ and $ k_{\rm topo}/k_0 $ we find two distinct regimes (Fig.~\ref{fig:figure2}a): the \textit{non-bursty} regime, identified by $\xi=0$, and the \textit{bursty} regime, for $ \xi>0 $ (a mean field theory gives a good prediction of the boundary between the two). In the most significant region ($k_{\rm topo}/k_0 \sim 1.4 $, $ \bar{J}/D \sim 1.5 $) the burst size is between $2$ and $3$, close to that measured in \textit{E. coli} \textit{in vivo}~\cite{Golding}. Estimates of the other bursts parameters -- burst duration and OFF-state duration -- are given in the Supporting Material, Section~III, and are also in good agreement with experimental results~\cite{Golding,Chong}. We note though that the burst size depends on the model parameters: the system can produce bursts of significantly more than $2$--$3$ events (at most $\sim 10$ on average in our simulations). However, this only occurs in the transition region between the non-bursty and bursty regimes (see Supporting Material, Fig.~S2a). In this transition region, burst significance is smaller, which means that the separation between timescales is less marked.

The results in Figure~\ref{fig:figure2}a also show that when the positive feedback between supercoil generation and transcription initiation is strong (for large $\bar{J}/D$ and $k_{\rm topo}=0$, identified as the {\it supercoiling-regulated regime} of Ref.~\cite{Brackley}), the dynamics are never bursty; bursts are most significant when this feedback is much weaker (but non-zero). The reason for this seemingly surprising result is that if supercoiling upregulates transcription too much, the gene is essentially always on and the transcriptionally silent state is absent (see Fig. S3c in Supporting Material).

Our results show that topoisomerases action favours burstiness. In other words, although the dynamics can be bursty for $k_{\rm topo}=0$, burst significance is larger when $k_{\rm topo}\ne 0$.

\begin{figure}[b]
\centering
{\includegraphics[width=.5\textwidth]{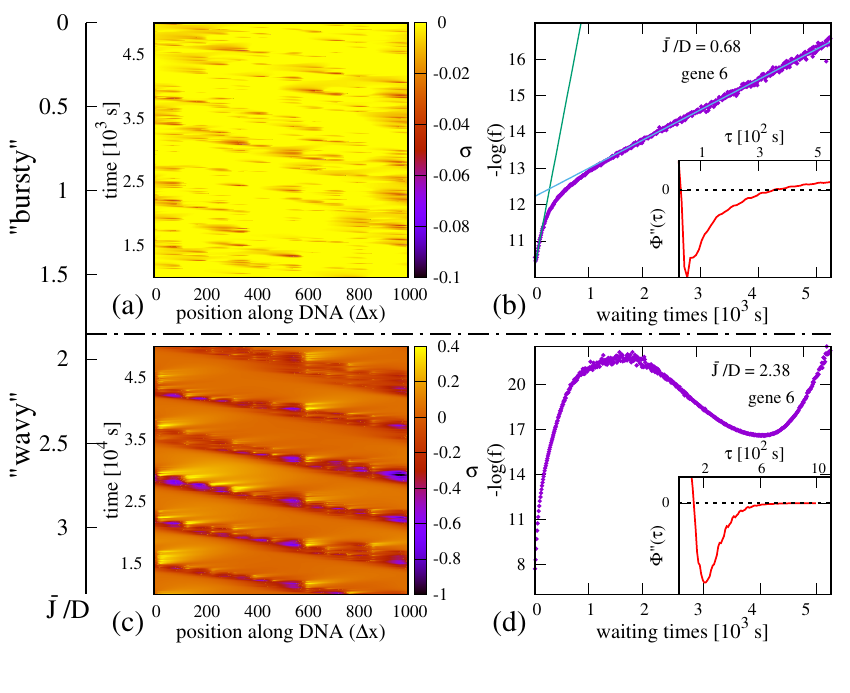}} \\
\caption{\textbf{Bursty and wavy regimes for an array of 10 tandem genes}. \textbf{(a)} Kymograph in the bursty regime ($ \bar{J}/D = 0.68 $). For clarity, we only show the negative supercoiling range. There are correlations between neighbouring genes, but no clear periodic pattern. Note that genes $1$ and $10$ turn red more often (compare to genomic map in Fig.~5a), since they are slightly upregulated. \textbf{(b)} Plot of the \textit{pdf} of waiting times in the bursty regime, showing the emergence of two timescales. Inset: second derivative of the \textit{ssf}, which displays two zeroes. \textbf{(c)} Kymograph in the supercoiling-regulated regime ($ \bar{J}/D = 2.38 $). For high values of the flux, the bursty dynamics are replaced by transcription waves. \textbf{(d)} \textit{Pdf} of waiting times in the supercoiling-regulated regime. The new timescale associated with the wave modifies the shape of the distribution, giving rise to a ``bump'' at $\sim4$$\times$$10^3$~s (local minimum in -log plot). Inset: second derivative of \textit{ssf}. In the physically relevant range of waiting times ($ \tau \lesssim 10^3 $~s) the function asymptomatically approaches $ 0 $ without crossing the axis ($ \tau_2 \rightarrow +\infty $).}
\label{fig:figure3}
\end{figure}

As bursting is generally due to switching back and forth between two transiently stable states, it is natural to ask whether there are any signatures of bistability in the stochastic transcriptional process we simulate. As we show in Fig.~\ref{fig:figure2}b, one such signature can be obtained from moments of the distribution of supercoiling at the promoter $ \sigma_p$. For non-bursty behaviour, $\sigma_p$ exhibits close-to-Gaussian fluctuations about an average value (see Supporting Material, Fig.~S3b). For bursty transcription, this distribution is more markedly non-Gaussian and bistable (see Supporting Material, Fig.~S3a). Quantitatively, burst significance correlates with the magnitude of non-Gaussianity parameters such as kurtosis and skewness (Fig.~\ref{fig:figure2}b, and Supporting Material, Fig.~S3). 

\subsection*{Multiple genes}

The single gene case considered above is an important starting point for our model, and could be relevant to the experimental investigation in Ref.~\cite{Golding} where the transcription of a gene on a bacterial plasmid was monitored. However, it is also of interest to consider the case of multiple genes. This is because gene density is variable both across organisms and within genomes: for instance, in both yeast and bacteria gene density is high so that transcription is likely to affect neighbouring genes. This is also relevant for understanding synthetic DNA constructs containing multiple genes, which can be used in biotechnology applications. In this Section, we study the burstiness in arrays of multiple genes without topoisomerases (i.e. $k_{topo} = 0$).

In Figure~\ref{fig:figure3} we consider the supercoiling-coupled transcriptional dynamics within an array of genes which have the same orientation (we refer to these as ``tandem'' genes). We find that bursts are typically more significant than in the single gene case. For instance, for $ \bar{J}/D = 1 $ and $k_{topo} = 0$, the single gene case was only weakly bursty ($\xi\sim 0.23$, burst size $\sim 1.53$ and duration $\sim 100$ s, see Fig.~S2 in the Supporting Material), whereas for an array of $10$ tandem genes the same parameters give rise to bursting which is around twice as significant ($\xi \sim 0.4$--$0.5 $ for the most bursty genes, burst size $\sim 2$ and duration $\sim 3$--$4$ min, see Fig.~S7 in the Supporting Material). This is because transcription generates positive supercoils ahead of a gene, which act to down-regulate its downstream -- right -- neighbour  (whilst upregulating the upstream -- left -- neighbour). As a result, some genes may be transiently ``switched off'' -- this can be appreciated, for instance, by inspecting the time series of supercoiling at the promoter, which at times can take sufficiently positive values such that $k_{\rm in}(t)=0$, see Fig.~S5a in the Supporting Material. Just as in the case of a single gene with $k_{\rm topo}\ne 0$, the activity of each gene is effectively described by a two-state dynamics (ON$\leftrightarrow$OFF), and bursts can occur (Fig.~\ref{fig:figure3}a). As expected, the \textit{pdf} of waiting times is well described by a double-exponential (see Fig.~\ref{fig:figure3}b), and $\Phi''(\tau)$ displays two zeros (see Inset Fig.~\ref{fig:figure3}b)

\begin{figure}[t]
\centering
{\includegraphics[width=.5\textwidth]{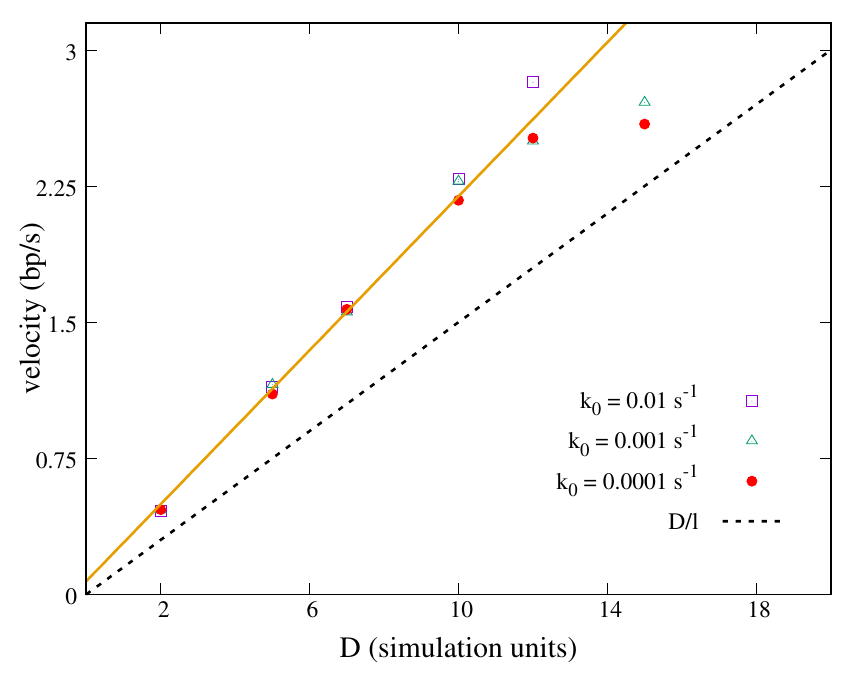}} \\
\caption{\textbf{Wave velocity and scaling relation.} Wave velocity for different values of $ D $ ($ \bar{J} = 25.5 $) and $ k_0 $, so that we span a large range of $\bar{J}/D$ values, between $ 1.7$--$12.7$. Values of $D$ are given in simulation units (i.e., in units of $\Delta x^2/\Delta t$). Simulation data are well fitted by a straight line (orange), whereas the simple scaling theory discussed in the text underestimates the data slightly (dashed black line).}
\label{fig:figure4}
\end{figure}

As for the single gene case, in the multi-gene system bursting does not occur for large values of $\bar{J}/D$. In this regime, the supercoiling-mediated intergenic interactions instead give rise to transcription waves which travel in the opposite direction to transcription (Fig.~\ref{fig:figure3}c,d). Transcription waves arise because transcription of a gene upregulates its upstream -- left -- neighbour~\cite{Brackley}: as a consequence, transcription of gene $i$ is followed by that of gene $i-1$, then $i-2$ and so on. We find that the wave velocity is $v\sim D/l$, independent of $k_0$ (Fig.~\ref{fig:figure4}, inset; $l = 100 \Delta x$ is the mean separation between promoters in our simulations).Given our parameter choice, the wave speed is between $0.5$--$3.0$ bp/s, and the time needed to trigger activity of the neighbouring upstream gene is between $6$--$12$ minutes. The scaling of $v$ can be understood by assuming that supercoiling propagates diffusively over the distance $l$ between a gene and its upstream neighbour -- simulations show that the prefactor in this relationship is slightly larger than $1$. When in the ``wavy'' regime, the system can no longer be mapped onto a telegraph-like process, and bursts are no longer observed (accordingly $\xi=0$, as $\Phi(t) $ does not have two inflection points, see Fig.~\ref{fig:figure3}d). 

\begin{figure}[t]
\centering
{\includegraphics[width=.5\textwidth]{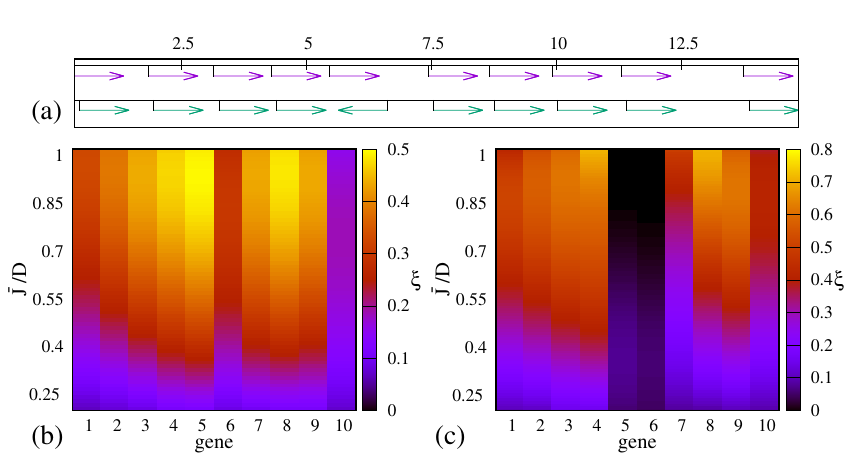}} \\
\caption{\textbf{Burstiness for a multiple genes arrays.}  \textbf{(a)} Map of gene positions for tandem (top) and divergent geometry (bottom). \textbf{(b,c)} Plot of $\xi$ for tandem geometry (b) and divergent geometry (c) (geometries used are in (a)). The range of $\bar{J}/D$ ($0.06$--$1.02$) is chosen so that the system is in the bursty regime.} 
\label{fig:figure5}
\end{figure}

\begin{figure}[t]
\centering
{\includegraphics[width=.5\textwidth]{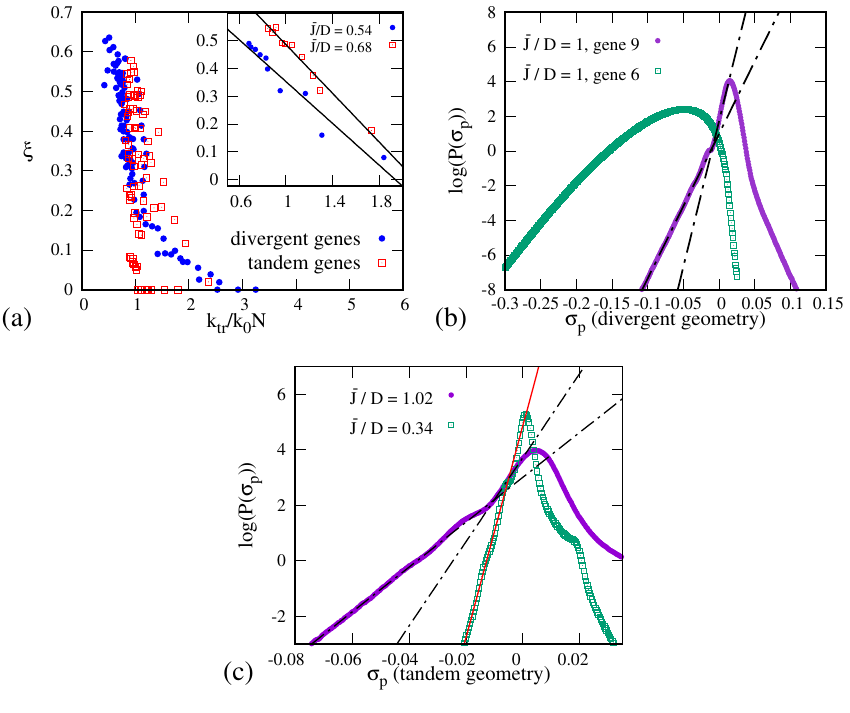}} \\
\caption{\textbf{Bursty and wavy regimes for an array of 10 tandem genes}. \textbf{(a)} The burstiness, $ \xi $, is plotted for different value of the overall time-averaged transcription rate $k_{\rm tr}$ for tandem (red squares) and bidirectional (blue circles) genes (each point represents a gene; points are shown for different values of $\bar{J}/D$ between $0.06$--$1.02$). Inset: for a given value of $ \bar{J}/D $, $ \xi $ depends linearly on $ k_{tr} $ (each point represents a gene). \textbf{(b)} Log-linear plot of the distribution of $\sigma_p$. For bursty genes (gene 9, which is not part of the divergent pair) this distribution shows a singularity or bump. Such a singularity does not appear for highly transcribed genes (gene 6, in the divergent pair). \textbf{(c)} For a given gene (gene 4, tandem geometry) the slope of the negative tail of the distribution depends on $ \bar{J}/D$. For small $ \xi $ (small $ \bar{J}/D $) the bump disappears.}
\label{fig:figure6}
\end{figure}

Transcription waves only arise for arrays of tandem genes, and do not occur (or do so only transiently) for genes of differing orientation. In that case transcription-generated supercoiling upregulates pairs of divergent genes at the expense of other (convergent or tandem) genes which are down-regulated~\cite{Brackley}: this renders the situation qualitatively closer to that of a single gene (see Supporting Material, Section~IV). For either kind of gene orientation, burstiness is gene-dependent, and the values of $\xi$ for different genes are substantially different from each other (Fig.~\ref{fig:figure5}). Note that even for small $\bar{J}/D$ some genes are slightly upregulated than others (as shown in the kymograph in Fig.~3a), due to the particular position in the array. We find that $\xi$ is anticorrelated with the overall transcription rate, so that the genes which are expressed more (e.g. genes $1$, $6$ and $10$ in the tandem setup or genes $5$ and $6$ in the divergent setup, see Fig.~5a and Supporting Material, respectively Figs.~S6 and S9a) are less bursty (Fig.~\ref{fig:figure6}a). Burst size is also different in the tandem and bidirectional setups, being substantially smaller in the latter case ($\beta \lesssim 2$, see Fig. S7a in the Supporting Material). Burst sizes measured experimentally in E. Coli~\cite{Golding} are closer to the value for single or tandem genes ($\beta\simeq 2.2$) -- this is reasonable, as they refer to the transcription of operons which are normally made up from tandem genes controlled by a single promoter.

An analysis of the distribution of supercoiling values at the promoter shows that these differ qualitatively for the cases of multiple genes and a single gene (e.g., compare Fig.~\ref{fig:figure6} with Fig.~S3).
Unlike in the single gene case, the non-Gaussian parameters for the distribution of supercoiling at the promoter now only weakly correlate with the burst significance (see Supporting Material, Fig.~S8). This is because the supercoiling-mediated interaction between genes give rise to non-Gaussian fluctuations even for non-bursty genes. Nevertheless, for both the tandem and divergent gene cases, bursting leaves a detectable signature in the tails of the distribution. 

For the bursty transcription case, there is a singularity or a bump, while for non-bursty transcription the curve is smooth, as shown in Figure~\ref{fig:figure6}b for a divergent geometry and in Figure~\ref{fig:figure6}c for a tandem array. The singular point is located at $ \sigma_p \approx (1-(k_0 \tau_x)^{-1})/\alpha $, where $ \tau_x = (\tau_1+\tau_2)/2$. We note that similar changes in the behaviour of large fluctuations were linked to phase transitions in other systems~\cite{Cagnetta,Janas}. 

\section*{Conclusions} 

 In summary, we have studied the occurrence of transcriptional bursts in a non-equilibrium model for supercoiling-regulated transcription, first introduced in Ref.~\cite{Brackley}. For an isolated gene, we found that significant bursting occurs primarily in the presence of topological enzymes which relax positive and negative supercoiling. This is qualitatively consistent with experimental evidence that bursts in {\it E. coli} arise due to the action of the DNA gyrase enzyme, which can relax positive supercoiling~\cite{Chong}. It is interesting to note that in the region of parameter space where bursts are most significant, the properties of the bursts generated in the model (size, duration and inter-burst time) match well those found in bacteria {\it in vivo}~\cite{Golding}. Notably, topoisomerase action is not required for highly significant bursting in gene clusters, as there supercoiling can mediate transient inhibition of the neighbours of highly active genes. We considered both tandem and bidirectional gene geometries, which could be recreated synthetically using plasmids of selected sequence in the presence of the transcriptional machinery. We found that the existence of bursting is intimately linked to the nature of fluctuations of supercoiling at gene promoters: bursting becomes most significant when these are strongly non-Gaussian. It would be of interest to look for such effects in experiments with populations of synthetic DNA loops, where psoralen binding~\cite{Naughton,Kouzine} might in principle be used to monitor averages and distributions of supercoiling along the DNA.

Whilst additional ingredients may be required to understand transcriptional bursts in eukaryotes, where stochastic promoter-enhancer interactions or other regulatory processes are known to play a key role~\cite{Corrigan,Raj,Fukaya}, our current results uncover a possible mechanism for transcriptional bursts in bacteria, based on the interplay between transcription initiation, supercoiling and topological enzymes. Our results are consistent with the work in Ref.~\cite{Sneppen}, which identified supercoiling as one potential mechanism for bursting based on comparison between experimental data and simplified kinetic model -- in our case, we consider a full stochastic dynamics for supercoiling and elucidate the role of topoisomerases.  
As discussed in the introduction, a related study~\cite{Levine2018} instead analysed the interplay between supercoiling and transcriptional {\it elongation}, providing a complementary mechanism to the one identified here; that model may be particularly relevant for the case of ribosomal genes where expression is extremely high, whereas the present work concerns more moderate expression where there is not significantly more than one polymerase transcribing the same gene at once. 

{\it Acknowledgements.} Simulations were run at Bari ReCaS e-Infrastructure funded by MIUR through PON Research and Competitiveness 2007–2013 Call 254 Action I. We acknowledge support from ERC (CoG 648050, THREEDCELLPHYSICS).

{\it Author  contributions.} MA, AB, CAB, GG, DM designed research; MA, AB performed simulations; MA analysed data; MA, AB, CAB, GG, DM wrote the paper.

\widetext
\clearpage
\begin{center}
\textbf{Transcriptional bursts in a non-equilibrium model for gene regulation by supercoiling -- Supplementary Information}
\end{center}

\setcounter{equation}{0}
\setcounter{figure}{0}
\setcounter{table}{0}
\setcounter{page}{1}
\makeatletter
\renewcommand{\theequation}{S\arabic{equation}}
\renewcommand{\thefigure}{S\arabic{figure}}
\renewcommand{\bibnumfmt}[1]{[S#1]}
\renewcommand{\citenumfont}[1]{#1}



\baselineskip=15pt

\definecolor{red}{rgb}{1,0,0}
\definecolor{blue}{rgb}{0,0,1}


\title{Transcriptional bursts in a non-equilibrium model for gene regulation by supercoiling \\ Supporting Material}

\author{M. Ancona}
\affiliation{SUPA, School of Physics and Astronomy, University of Edinburgh, Peter Guthrie Tait Road, Edinburgh, EH9 3FD, UK}
\author{A. Bentivoglio}
\affiliation{SUPA, School of Physics and Astronomy, University of Edinburgh, Peter Guthrie Tait Road, Edinburgh, EH9 3FD, UK} 
\author{C.~A. Brackley}
\affiliation{SUPA, School of Physics and Astronomy, University of Edinburgh, Peter Guthrie Tait Road, Edinburgh, EH9 3FD, UK}
\author{G. Gonnella}
\affiliation{Dipartimento di Fisica, Universit\`a di Bari and INFN, Sezione di Bari, 70126 Bari, Italy}
\author{D. Marenduzzo}
\affiliation{SUPA, School of Physics and Astronomy, University of Edinburgh, Peter Guthrie Tait Road, Edinburgh, EH9 3FD, UK}

\maketitle

\section{MEAN FIELD THEORY}
\label{sec:mean field}

\renewcommand{\theequation}{S\arabic{equation}}
\renewcommand{\thefigure}{S\arabic{figure}}

Here, we develop a mean field theory with some improvements with respect to our previous work~\cite{Brackley}. In particular we solve the mean field ordinary differential equation (ODE) with periodic boundary conditions (instead of open boundary conditions, as previously done) and in the presence of topoisomerases. 

We consider the case $ N = n = 1 $, where $N$ is the number of RNAP and $n$ is the number of genes. Besides, we consider a \textit{static} polymerase (i.e. $v$ = 0) at the lattice position $x=0$. If $ L $ is the length of the lattice, we assume boundary conditions $ \sigma(0) = 0 $ and $ \sigma(L/2) = \sigma(-L/2) $. In steady state ($ \partial \sigma / \partial t = 0$), Eq. (1) reads:

\begin{equation}
\frac{\partial^2 \sigma(x)}{\partial x^2} - \frac{J_0}{D} \frac{k_{\rm in} \tau}{k_{\rm in} \tau + 1} \frac{\partial \delta(x)}{\partial x} - \frac{k_{\rm topo}}{D} \sigma(x) = 0
\label{seq:meanfield1}
\end{equation}
where we have made the mean field approximation 
\begin{equation}
\dfrac{J_{tr}(x,t)}{D} \rightarrow \dfrac{J_0}{D} \frac{k_{\rm in} \tau}{k_{\rm in} \tau + 1} \delta(x)  \equiv M \delta(x)
\label{seq:flux}
\end{equation}
with $ k_{\rm in} \tau/(k_{\rm in} \tau + 1) $ the fraction of time the system spends in the transcribing state. 

As the flux term acts only at $ x = 0 $, solving the model in the mean field approximation is equivalent to solving the following ODE:
\begin{equation}
\begin{cases}
\dfrac{\partial^2 \sigma(x)}{\partial x^2} - \dfrac{k_{\rm topo}}{D} \sigma(x) = 0 \qquad \qquad x \neq 0 \\ \\
\dfrac{\partial \sigma(x)}{\partial x}\bigg|_{x=0} = M \delta(0) \\ \\
\sigma\left(L/2\right) = \sigma\left(-L/2\right).
\end{cases}
\label{seq:meanfield2}
\end{equation}

Since both $ \sigma(x) $ and $ \sigma(-x) $ are solution of the ODE for $x\neq 0$ the unique solution of Eq.~\eqref{seq:meanfield2} is a linear combination of $\sigma(x)$ and $\sigma(-x)$. It can be shown that only the antisymmetric combination fulfils the periodic boundary conditions, with $ \sigma(L/2) = \sigma(-L/2) = 0 $. 



The solution of Eq.~\eqref{seq:meanfield2} with the appropriate parity and boundary conditions is given by:
\begin{equation}
\sigma(x) = \dfrac{M}{2} \dfrac{{\rm sinh}\left[\sqrt{\dfrac{k_{\rm topo}}{D}} \left(\dfrac{L}{2}-|x|\right)\right]}{{\rm sinh}\left[\sqrt{\dfrac{k_{\rm topo}}{D}} \dfrac{L}{2}\right]} {\rm sgn}(x),
\label{seq:meanfield4}
\end{equation}
where sgn(x) is the \textit{sign function}. From Eq.~\eqref{seq:meanfield4} it can be easily shown that in the limit $ k_{\rm topo} \rightarrow 0 $ we obtain:
\begin{equation}
\sigma(x) = \dfrac{M}{2} \left(1-\dfrac{2|x|}{L}\right) {\rm sgn}(x).
\label{seq:meanfield5}
\end{equation}
The term proportional to $ 1/L $ is the correction due to the periodic boundary conditions, that disappears for $ L \rightarrow \infty $, recovering the solution in Ref.~\cite{Brackley}. 
In the limit $ L \rightarrow \infty $, with finite $ k_{\rm topo} $, we have
\begin{equation}
\sigma(x) = \dfrac{M}{2} \exp\left(-\sqrt{\dfrac{k_{\rm topo}}{D}}|x|\right) {\rm sgn}(x).
\label{seq:meanfield5}
\end{equation}

The validity of this mean field theory can be determined by comparing it to the time-average supercoiling profile in our single gene simulations. 

\if{At this point, we can find a self-consistent equation for the transcriptional rate $ k_{on} $, by using Eq.~\eqref{seq:rate}. Using the same procedure shown in \cite{Brackley,Bentivoglio}, and setting $ k_{\rm topo}/D \equiv \omega $, we find
\begin{equation}
k_{on} = \dfrac{h(x) + \sqrt{h^2(x) + 4k_0\tau}}{2\tau}
\label{eq:kon}
\end{equation}
where
\begin{equation}
h(x) = k_0\tau\left(1 + \dfrac{\alpha J_0}{2D} \dfrac{{\rm sinh}\left[\sqrt{\omega} (L/2 - |x|) \right]}{{\rm sinh}\left[\sqrt{\omega} L/2 \right]}\right) - 1.
\label{seq:h}
\end{equation}

Still, we find the crossover by imposing $ h(x) \sim 0 $, that corresponds to that region in the parameters space where $ k_{on} $ starts to be significantly affected by the supercoiling; this occurs when ($ |x| \equiv -x_j $)

\begin{equation}
 2 \cdot \dfrac{sinh\left[\sqrt{\omega} L/2 \right]}{sinh\left[\sqrt{\omega} (L/2 + x_j) \right]} = \dfrac{J_0}{D}.
\label{seq:crossover}
\end{equation}

Posing $ A \equiv L/2 $, $ B \equiv A + x_j $, if $ \sqrt{\omega} A < \sqrt{\omega} B \ll 1 $, we can expand \eqref{seq:crossover}, and we obtain

\begin{equation}
\dfrac{J_0}{D} = 2 \cdot \dfrac{A + A^3\omega + o(\omega^{2})}{B + B^3\omega + o(\omega^2)} \simeq \dfrac{2}{\left(1+\dfrac{2x_j}{L}\right)} + \dfrac{L^2x_j}{2} \omega.
\label{seq:firstlimit}
\end{equation}
 We remark that for $ \omega = 0 $ and $ L \rightarrow +\infty $ we find $ J_0/D \sim 2 $. 

Conversely, if $ L \rightarrow \infty $, we should rewrite $ h(x) $ using \eqref{seq:meanfield5}. We find

\begin{equation}
h(x) = k_0\tau\left[1 + \dfrac{\alpha J_0}{2D} exp( \sqrt{\omega} x_j ) \right] - 1.
\label{seq:h2}
\end{equation}
In the limit $ \omega \ll 1 $, we have 

\begin{equation}
\dfrac{J_0}{D} \simeq 2 + 2\sqrt{\omega} x_j 
\label{seq:h2}
\end{equation}
}\fi
\vspace{3mm}

Interestingly, from our simulations we found that the point along the gene at which the time-averaged supercoiling profile crosses zero is $\sim 2\lambda/3 $, independently of the parameter used. The correct mean field profile of supercoiling for a moving polymerase is then computed by substituting $ |x| \rightarrow |x - 2\lambda/3 |$.
\section{THE MODIFIED \textit{IPP} PROCESS}

The mechanism which leads to bursty dynamics for transcription in living cells is still not well understood, though several hypothesis have been made. In our model, we have seen that both the action of topoisomerases (1-gene model) and the interaction among genes (10-gene model) can yield bursts, in absence of external factors. The nontrivial nonlinear behaviour predicted by our model can be captured by a simpler kinetic scheme: the \textit{Interrupted Poisson Process} (\textit{IPP}). By solving the \textit{IPP} equations, one can explicitly obtain the double-exponential distribution of waiting times between events, that, when appropriate conditions on the kinetics rates are met, leads to bursting. In this section we modify the \textit{IPP} equations. Nevertheless, the resulting distribution is still a double-exponential (see below), with the exception of the two timescales, which are different from those found in~\cite{Dobrzynsky}.
\vspace{5mm}

We define (as in Fig.~1a of the main text) an \textit{active}  ($ ON $, or $\epsilon=1$) and an \textit{inactive} ($ OFF $, or $\epsilon=0$) state of the gene promoter according to the local supercoiling density (i.e., if the supercoiling density is below $(1-(k_0 \tau_2)^{-1})/\alpha$, then the promoter is $ ON $).
 We then associate the rates $ k_{OFF} $ and $ k_{ON} $  with the $ ON \rightarrow OFF $ and $ OFF \rightarrow ON $ transitions respectively. The gene oscillates between the two states, tracking the typical trajectories of a \textit{Random Telegraph Process} (see Fig.~1a, BOTTOM). Whilst in the $ ON $ state, the gene is able to transcribe with rate $ k_i $. Given a time series of events, the \textit{waiting time} $ t_n $ is the elapsed time between two consecutive transcriptions, say the $ (n-1) $-th and the $ n $-th. In the \textit{IPP}, the waiting time $ t_n $ is drawn by the same probability distribution function (\textit{pdf}) for each $n$ (this may not be true in general in our stochastic model for supercoiling-dependent transcription). Moreover, in the standard \textit{IPP} prescription the $ OFF \rightarrow ON $ transition occurs between states labelled by the same $n$, see Ref.~\cite{Dobrzynsky}.
Differently, in our modified \textit{IPP} description, the transition $ OFF \rightarrow ON $ is triggered by transcription (see Fig.~\ref{fig:schema_cinetico1}), meaning that the transition and the first event in the burst occur at the same time.
By labelling possible states with the instantaneous value of $\epsilon$ (gene ON/OFF) and by the index $n$, which keeps count of the number of transcriptions, we denote the probability of being in the state $ \{n,\epsilon\} $ at time $t$ by $ p_{n,\epsilon}(t) $. Then, the set of master equations for our modified \textit{IPP} is:

\begin{figure}[t]
\centering
\includegraphics[width=.8\textwidth]{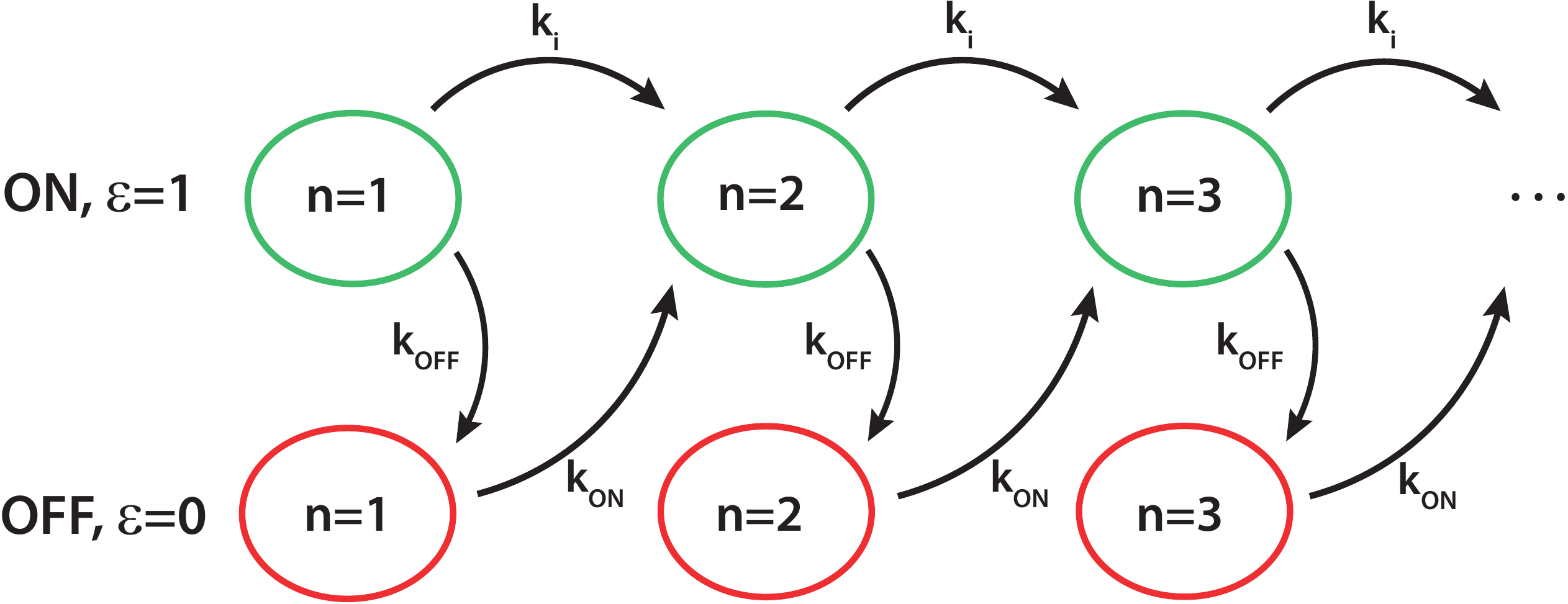}
\caption{\textbf{Scheme of the discrete states in a \textit{modified IPP}.} Each \textit{box} represent a state, with $ \epsilon \in \{0,1\} $. The index $ n $ labels the number of initiation events.}
\label{fig:schema_cinetico1}
\end{figure}

\begin{equation}
\begin{cases}
\dfrac{dp_{1,1}(t)}{dt} = -(k_i + k_{OFF}) \ p_{1,1}(t) \\
\dfrac{dp_{n,1}(t)}{dt} =  k_i p_{n-1,1}(t) + k_{ON} p_{n-1,0}(t) - (k_i + k_{OFF}) p_{n,1}(t), \qquad n = 2,3,\ldots \\
\dfrac{dp_{n,0}(t)}{dt} = k_{OFF} \ p_{n,1}(t) - k_{ON} \ p_{n,0}(t) \qquad n = 1,2,\ldots
\end{cases}
\label{eq:masterEq1}
\end{equation}
with the initial condition $ p_{1,1}(t = 0) = 1 $. Clearly, the \textit{pdf} associated with the first transcription event after initialisation ($n=2$) corresponds to the distribution of waiting times, that is
\begin{equation}
f(t) = k_i \ p_{1,1}(t) + k_{ON} \ p_{1,0}(t).
\label{eq:wt}
\end{equation} 
In order to find $ p_{1,1}(t) $ and $ p_{1,0}(t) $, and therefore $ f(t) $, we need to solve just the following two first order coupled ODEs:  
\begin{equation}
\begin{cases}
\dfrac{dp_{1,1}(t)}{dt} = -(k_i + k_{OFF}) \ p_{1,1}(t), \\
\dfrac{dp_{1,0}(t)}{dt} = k_{OFF} \ p_{1,1}(t) - k_{ON} \ p_{1,0}(t).
\end{cases}
\label{eq:sopravvivenza2}
\end{equation}  
By solving Eq.~\eqref{eq:sopravvivenza2} and using Eq.~\eqref{eq:wt}, we have

\begin{equation}
f(t) = w_1 r_1 e^{-r_1 t} + w_2 r_2 e^{-r_2 t},
\label{eq:soluzione1}
\end{equation}  
with rates $ r_{1,2} $ 
\begin{equation}
r_1 = k_i + k_{OFF}, \qquad r_2 = k_{ON},
\label{eq:rate1}
\end{equation}
and weights $ w_{1,2} $
\begin{equation}
w_1 = \frac{k_i - r_2}{r_1 - r_2}, \qquad w_1 \in \left[0,1\right],
\label{eq:peso1}
\end{equation}
\begin{equation}
w_2 = 1 - w_1.
\label{eq:peso2}
\end{equation}

In our stochastic model used in the main text, we can identify $ k_{ON} \sim k_0 $. Conversely, it is not easy to find a value for the rate $ k_{OFF} $ without fitting the data, since the transition $ ON \rightarrow OFF $ is mainly due to fluctuations, that, in the bursty phase, relax the system towards the initial value of the supercoiling $ \sigma_0 $.

\section{1-GENE ARRAY: BURST PARAMETERS AND ADDITIONAL FIGURES}

\begin{figure}[b]
\centering
\captionsetup[subfloat]{labelfont=bf}
\subfloat[][]{\includegraphics[width=.49\textwidth]{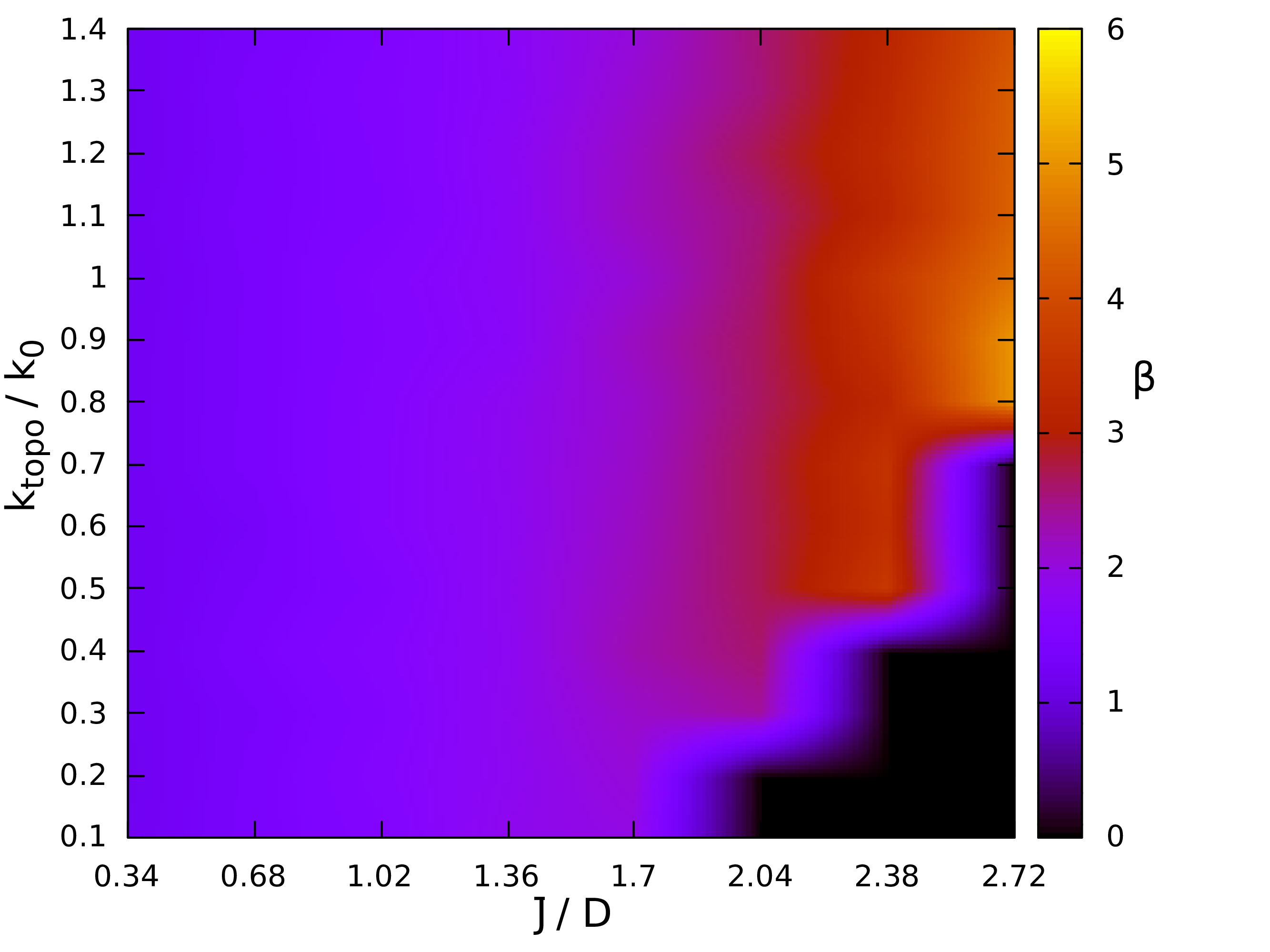}}
\subfloat[][]{\includegraphics[width=.49\textwidth]{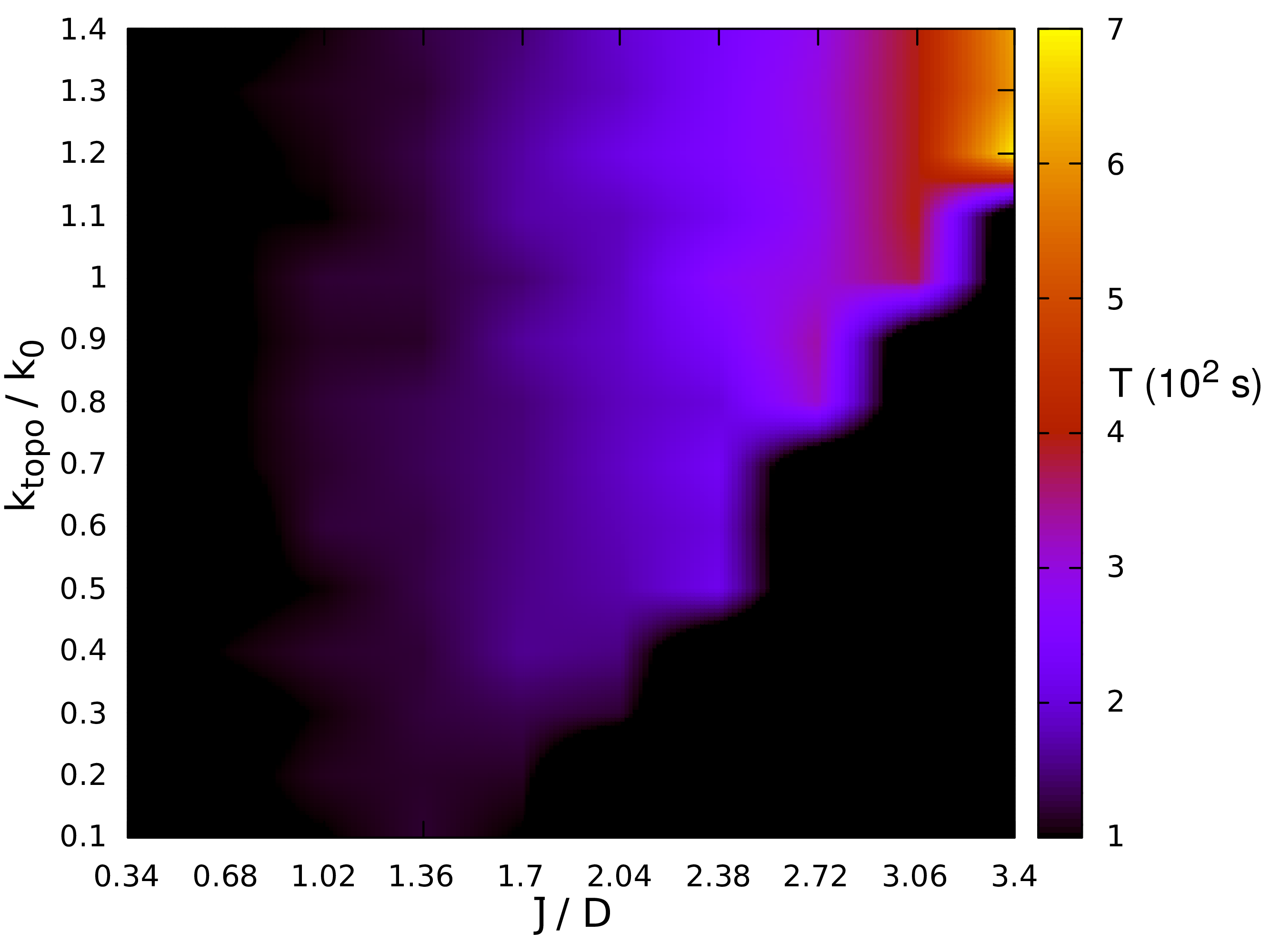}}
\caption{\textbf{Burst size and burst duration for a single gene}. \textbf{(a)} Here, we show the burst size $\beta$ up to $ \bar{J}/D = 2.72 $, in order to highlight the region of higher burst significance ($ \bar{J}/D \sim 1.5-2 $), in which $ \beta $ is in agreement with experimental measurement of the same parameter \cite{Golding}. \textbf{(b)} Duration of bursts $T$. We find that in the same region the duration of bursts is also consistent with \cite{Golding}, as $ T \sim 3-4 $ min.}
\label{sfig:figS1}
\end{figure}

In our work we use the \textit{sequence-size function}, $\Phi(\tau)$, to analyse burst significance. Although this method has been presented in previous works~\cite{Dobrzynsky,Kumar}, it has not previously been applied to simulation of the dynamics of transcription which does not use predetermined kinetic rates for the process. 

\if{We originally used the significancy parameters defined in \cite{Dobrzynsky}:
\begin{equation}
\xi' = \dfrac{\tau_2 - \tau_1}{\tau_2}
\label{seq:burst_significance_prev}
\end{equation}
where $ \tau_{1,2} $ are the two roots of $ \Phi''(\tau) $. 

This parameter distinguishes the two phases -- bursty versus non-bursty -- since it equals $0$ by definition if one of the two roots does not exist (and if this is the case, the dynamics is never bursty). Nevertheless, in our model it is ineffective in measuring the burst significance in the bursty phase, as it is almost uniform in that phase, see Fig.~\ref{sfig:figS1}a.}\fi

We use the parameter $\xi$ as defined in Eq.~(7) in the main tex, which is different from the parameter proposed in \cite{Dobrzynsky}, $(\tau_2 - \tau_1)/\tau_2$. Indeed, the former parameter does show variation within the bursty phase, whereas the latter does not. Our parameter $\xi$ yields a reasonable estimate of burst significance, as (i) it is still proportional to $ \tau_2 - \tau_1 $ and (ii) it fulfils the intuitive expectation that the burst significance should decrease if the system spends more time in intermediate states, at fixed $ \tau_2 - \tau_1 $.

The analysis of the \textit{ssf} allows us to readily compute other relevant burst parameters in a relatively simple way. The time separation between the two timescales is just $ \tau_x = (\tau_1 + \tau_2)/2 $ and, from the definition of the \textit{ssf}, we can estimate the \textit{mean burst size} (the average number of transcriptions in a single burst) as $ \beta = \Phi(\tau_x) $. This is a useful parameter, since it provides a simple basis to compare with experimental data. As we can see from Fig.~\ref{sfig:figS1} higher values of $ \beta $ ($ \beta > 4-5 $) correspond to less significant bursting (see main text, Fig.~2). Conversely, in the region of higher $ \xi $ ($ \bar{J}/D \sim 1.5-2 $), the burst size $ \beta \simeq 2.2 $ is short and close to that experimentally observed in E. Coli~\cite{Golding}. The burst \textit{duration} -- i.e. the time duration of a single burst -- is estimated by $ \beta \tau_x $: in the same parameter region it is also consistent with experimental results~\cite{Golding,Chong}, $ T \sim 3-4 $ min.
\vspace{5mm}

\begin{figure}[bp]
\centering
\captionsetup[subfloat]{labelfont=bf}
\subfloat[][]{\includegraphics[width=.42\textwidth]{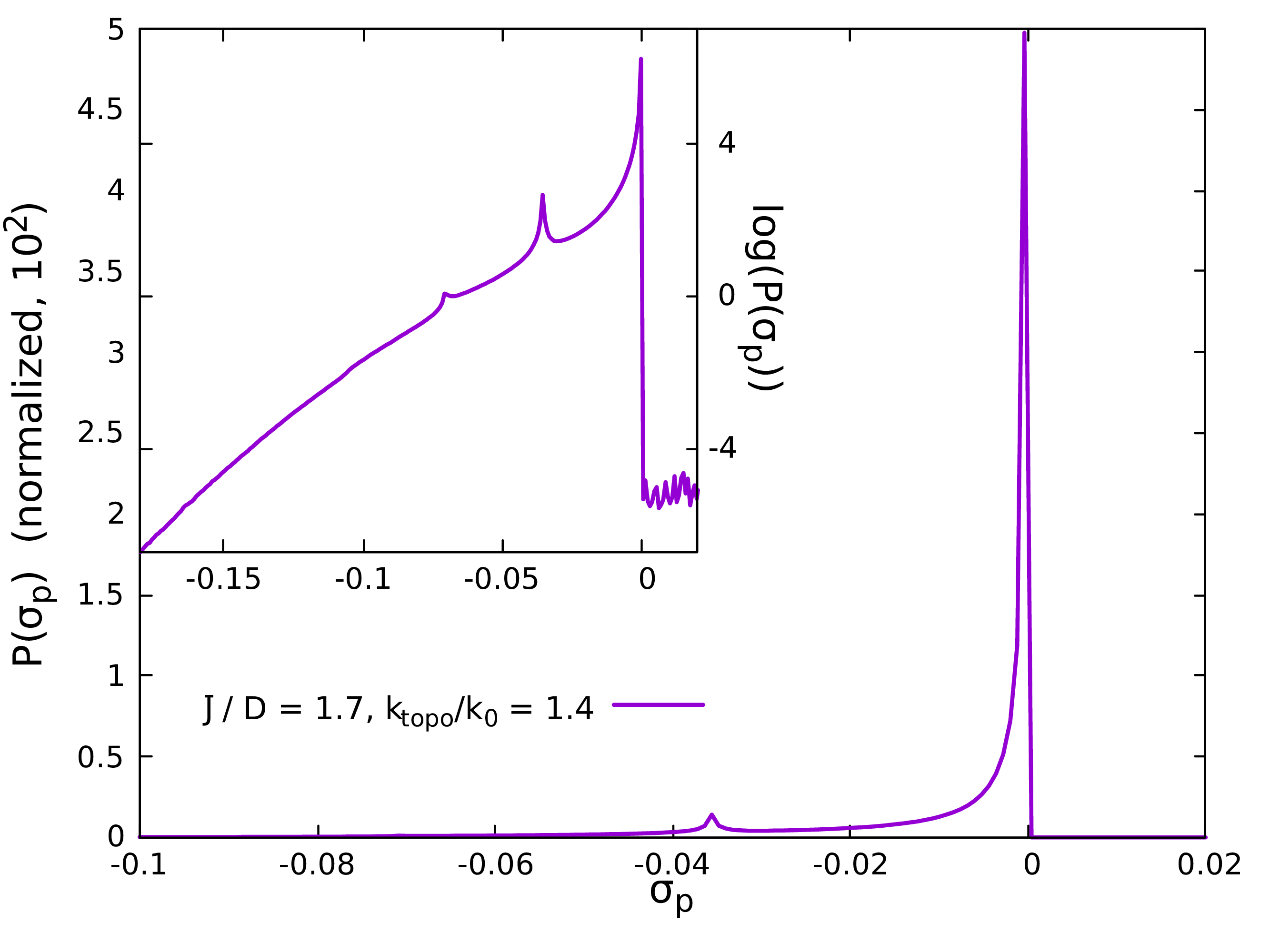}}
\subfloat[][]{\includegraphics[width=.42\textwidth]{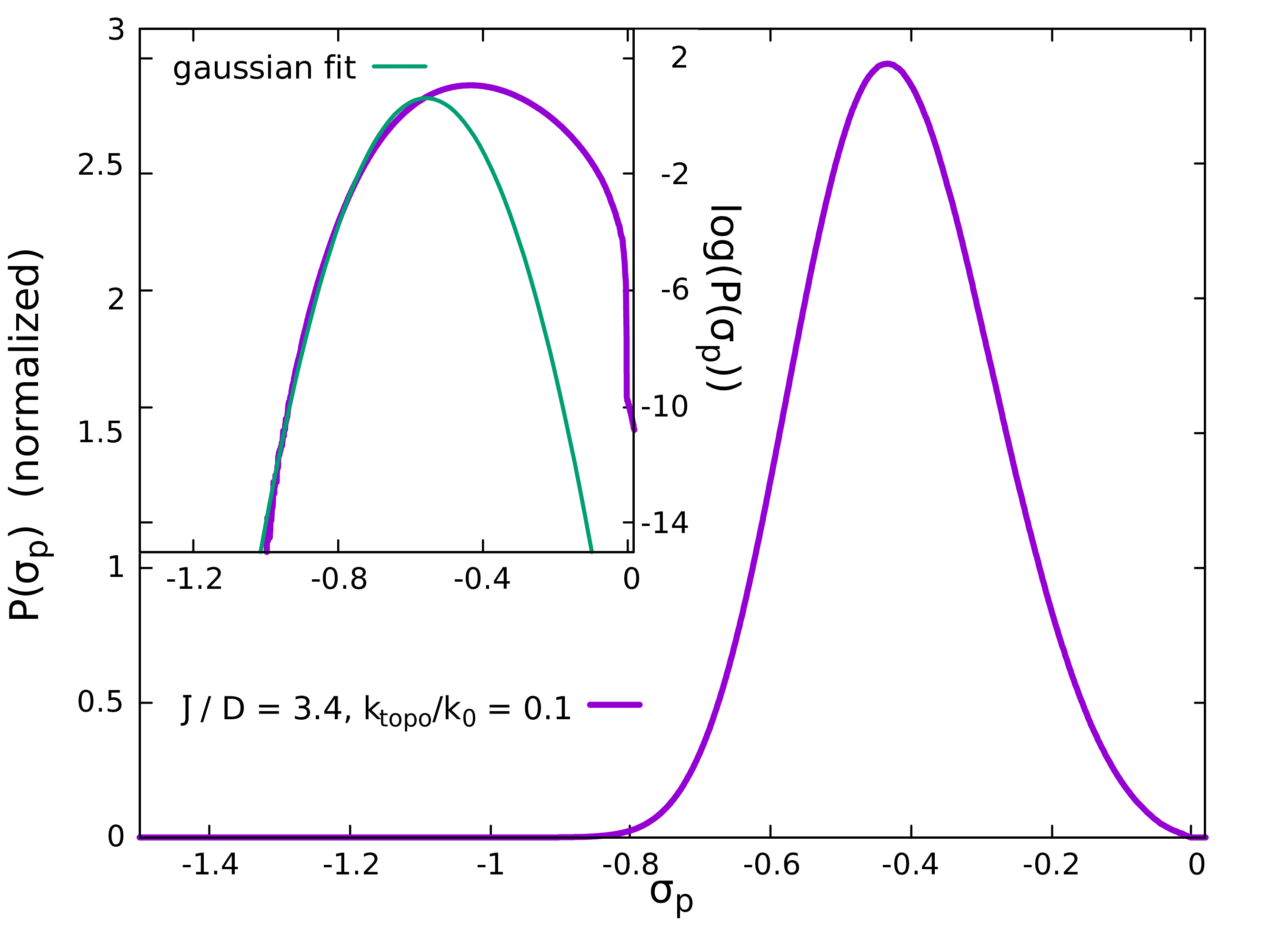}} \\
\subfloat[][]{\includegraphics[width=0.85\textwidth]{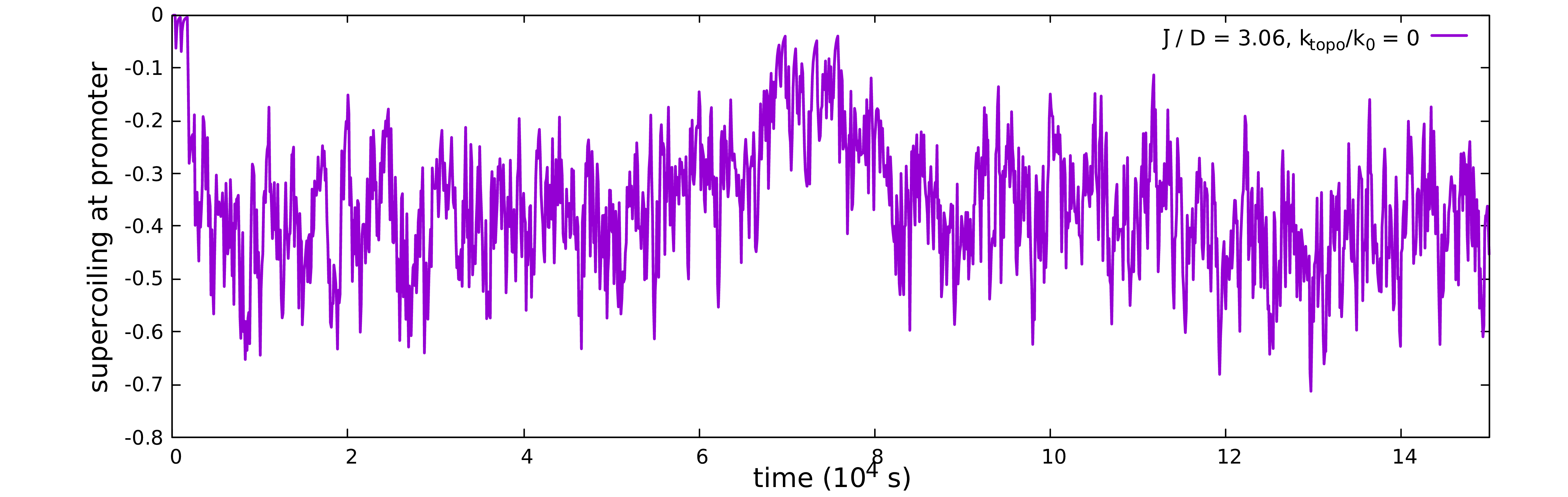}} 
\caption{\textbf{Probability distribution for supercoiling at the promoter}. \textbf{(a)} In the bursty phase, supercoiling at the promoter is strongly peaked at $ \sigma_p \sim 0 $. Another peak appears for more negative value of supercoiling, due to occupation of the ON state. Inset: log-linear plot of the \textit{pdf} in the main panel. \textbf{(b)} In the non-bursty phase the distribution is unimodal, with one gaussian tail. The gene tends to more often be in a state with less negative supercoiling for longer time; this results in a non-gaussian positive tail, with a nonzero kurtosis (see main text). \textbf{(c)} Time profile of the supercoiling at the promoter in the non-bursty regime. Clearly, the gene is always $ON$, as the supercoiling does not relax to the initial value $ \sigma_0 = 0 $.}
\label{sfig:figS3}
\end{figure}

In Fig.~\ref{sfig:figS3}a,b we present the typical probability distribution of supercoiling at the promoter, respectively in the bursty and non-bursty phases. Within our stochastic model, the supercoiling at the promoter is directly linked to the probability of initiation, and therefore its distribution encodes all of the information about the process. As expected, for bursty dynamics we observe a bimodal distribution of $ \sigma_p $, while for non-bursty dynamics we have a unimodal distribution, with fluctuations approximately Gaussian. 

\begin{figure}[t]
\centering
\captionsetup[subfloat]{labelfont=bf}
\subfloat[][]{\includegraphics[width=.4\textwidth]{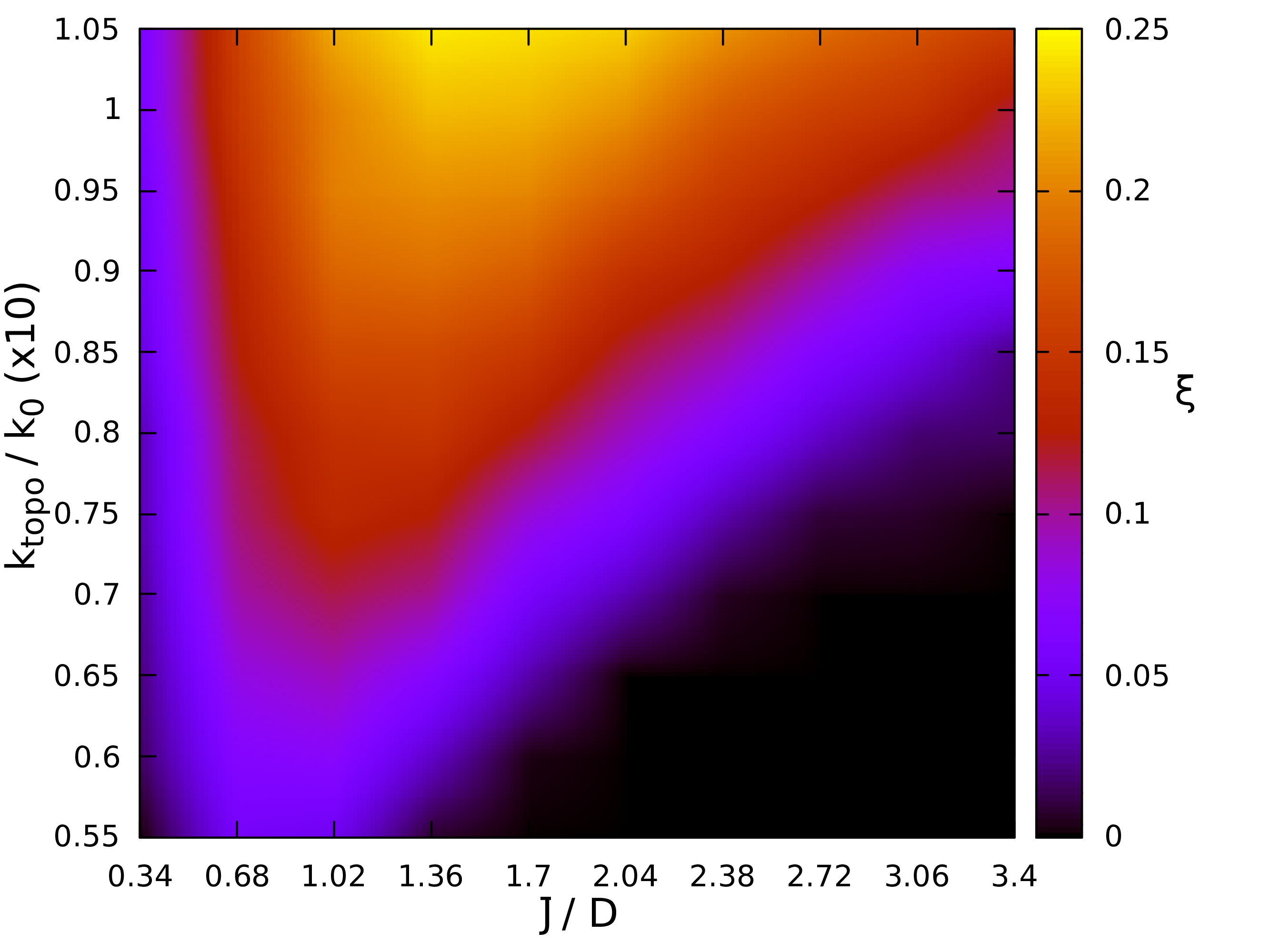}}
\subfloat[][]{\includegraphics[width=.4\textwidth]{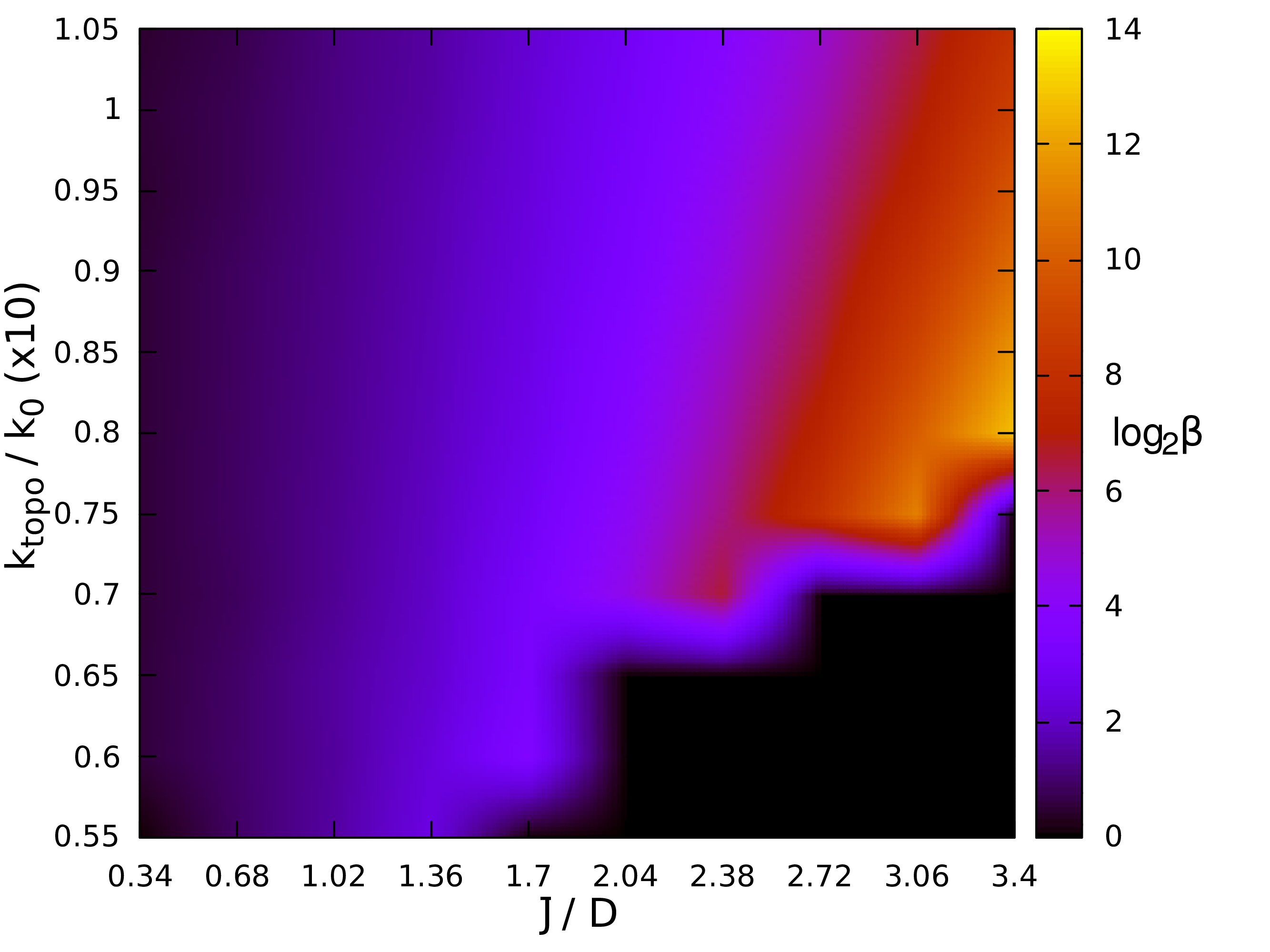}} \\
\subfloat[][]{\includegraphics[width=.4\textwidth]{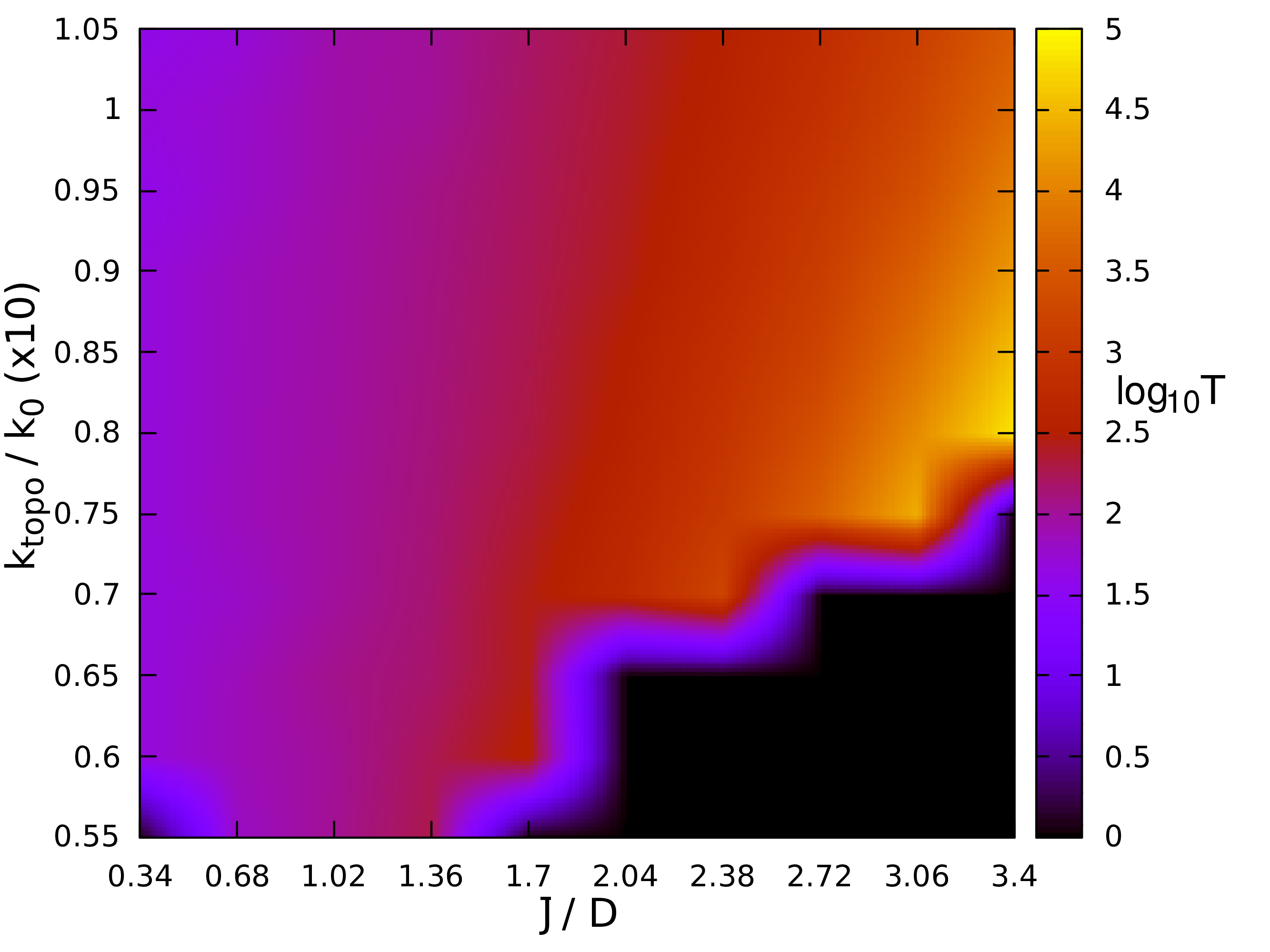}}
\caption{\textbf{Burst parameters for a single gene and three polymerases}. \textbf{(a)} Burst significance $\xi$. There are still two phases separated by a crossover. \textbf{(b,c)} Burst size $\beta$ and bursts duration $T$. With respect to the single gene case, here we have a large region of very high values (not biologically relevant) for both the burst size and the duration. 
However, these correspond to a region of low burst significance; in the region where burst significance is maximal we again find values for $ \beta $ and $ T $ consistent with experiments ($ \beta \sim 3 $, $ T \sim 2 $ min).}
\label{sfig:figS4}
\end{figure}

\vspace{5mm}
For completeness, in Fig.~\ref{sfig:figS4} we consider a situation where there is a single gene but three polymerases. The rationale is that \textit{in vivo}, at any given time, there can be more than one polymerase available for a given gene, even if the ratio of the total number of RNAP and genes is small. 
In this multiple polymerase case we find qualitatively similar results to the single polymerase case treated in the text, but only if we increase $ k_{\rm topo} $ by a factor of 10. However, the burst significance is remarkably smaller than the case studied in the main text. Nevertheless, the physical features of the bursts are consistent with those of the single polymerase case: e.g., for $ \bar{J}/D \sim 1 $ we find $ \beta \sim 3 $ and $ T \sim 2 $ min.

\clearpage

\section{MULTIPLE GENE SIMULATIONS: ADDITIONAL FIGURES}

We present some additional results from the $10$-gene array simulations for the case $k_{topo} = 0$, for which the main results are presented in the main text. 

In the case of tandem genes, for the configuration shown in the main text (see Fig.~5a) the genes $1$, $6$ and $10$ are upregulated by supercoiling. These genes have a larger space upstream of them, so are less affected by the repressive action of positive supercoils generated at their upstream neighbour. This occurs, albeit to a much lesser extent, even in the relaxed regime. This relatively small upregulation is sufficient to yield a sizeable change in the burst significance (see main text, Fig.~5b).

\begin{figure}[h]
\centering
\captionsetup[subfloat]{labelfont=bf}
\subfloat[][]{\includegraphics[width=.48\textwidth]{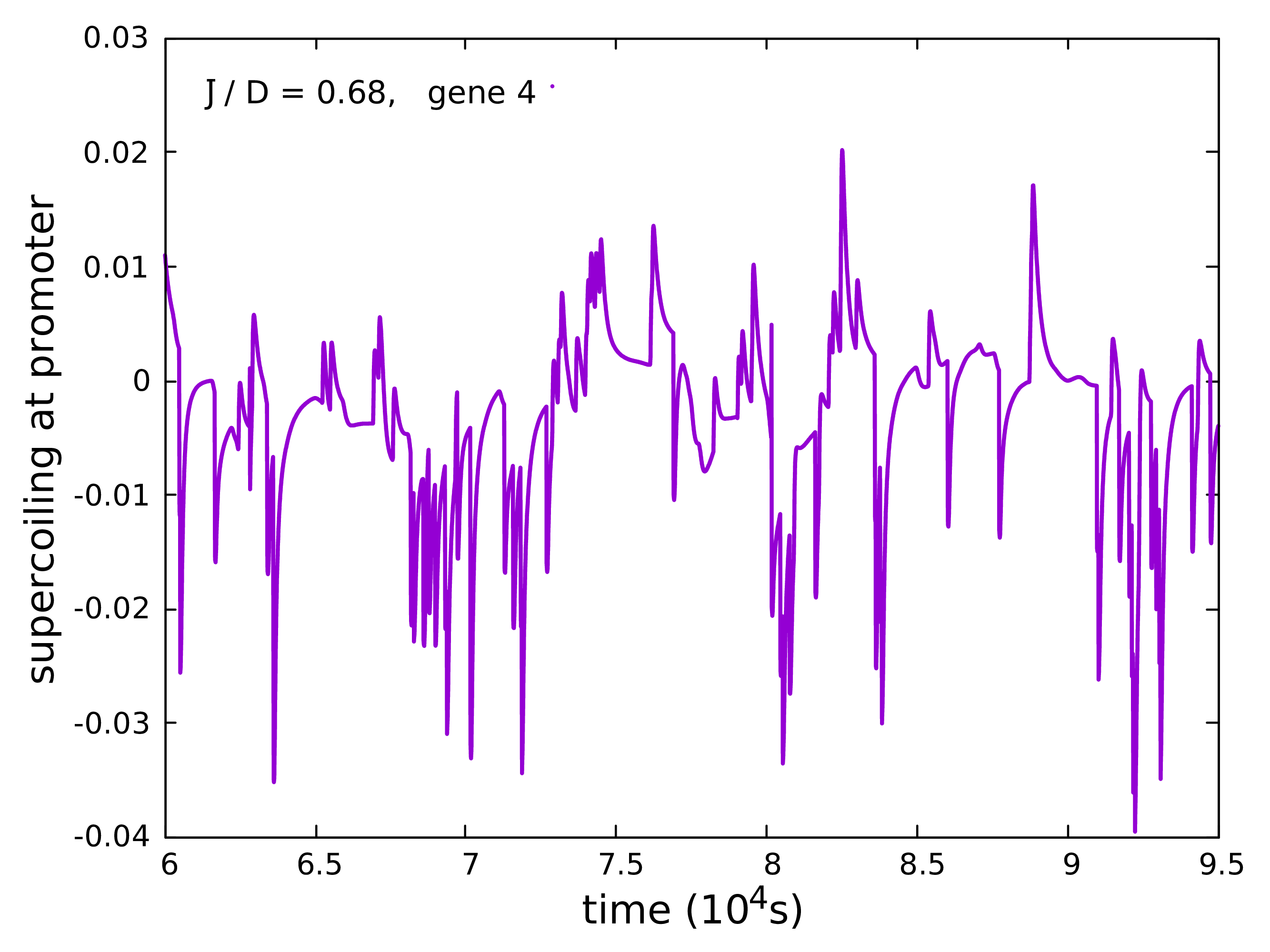}}
\subfloat[][]{\includegraphics[width=.48\textwidth]{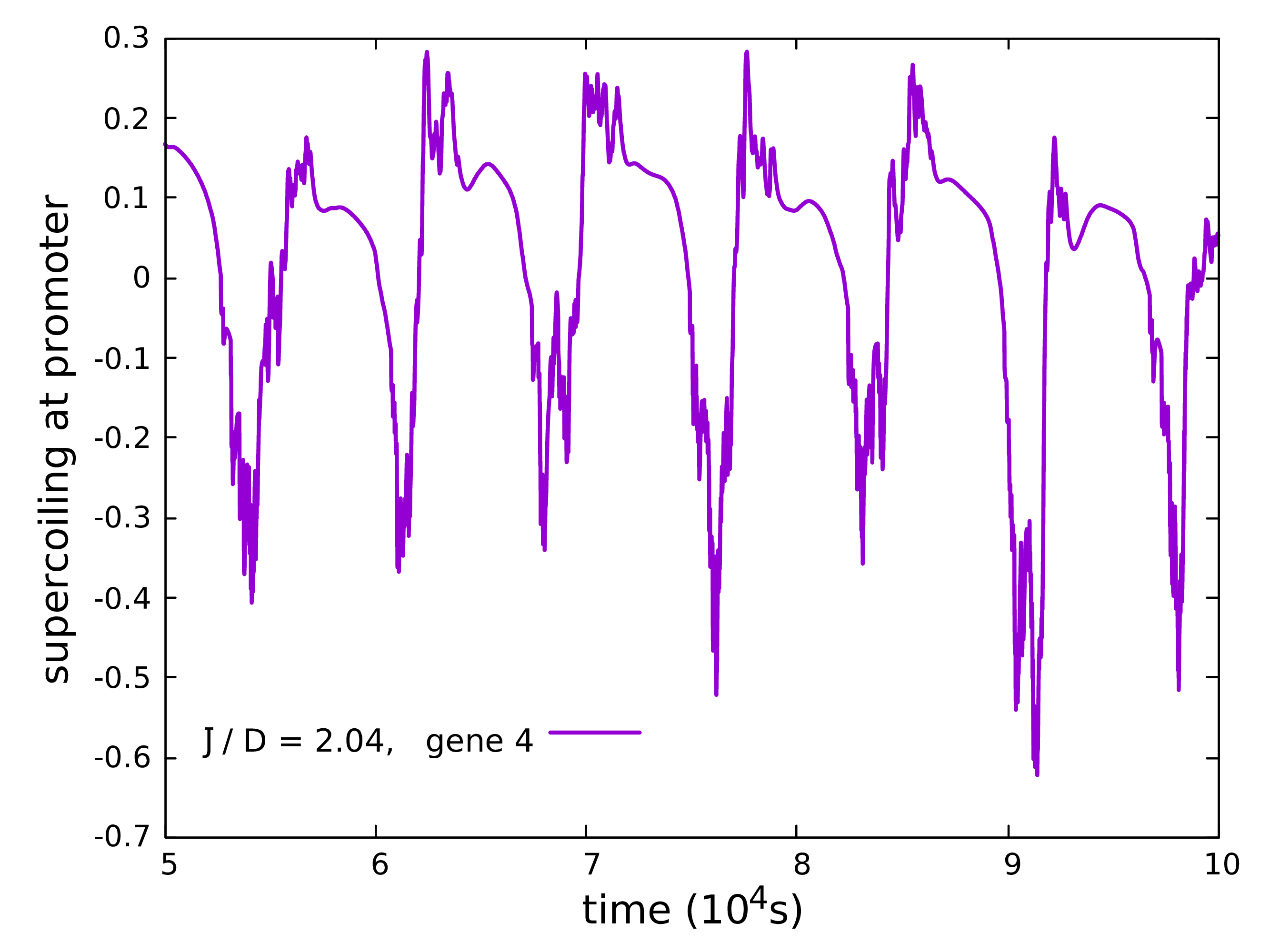}} 
\caption{\textbf{Supercoiling dynamics at the promoter gene 4 in a 10-gene array}. \textbf{(a)} Promoter supercoiling versus time in the bursty regime. For $ \bar{J}/D $ sufficiently small correlation between neighbours genes are established. Positive supercoiling produced by gene 3 transcription often freezes gene 4, yielding the supercoiling value transiently to the absorbing state ($ \sigma_0 = 0.01 $, $k_{in} = 0$). \textbf{(b)} Promoter supercoiling versus time in the supercoiling-regulated regime. In this regime the correlation spreads through the whole lattice, creating a transcription wave. Supercoiling at the promoter now oscillates in time.}
\label{sfig:figS5}
\end{figure}

\begin{figure}[bp]
\centering
\includegraphics[width=.48\textwidth]{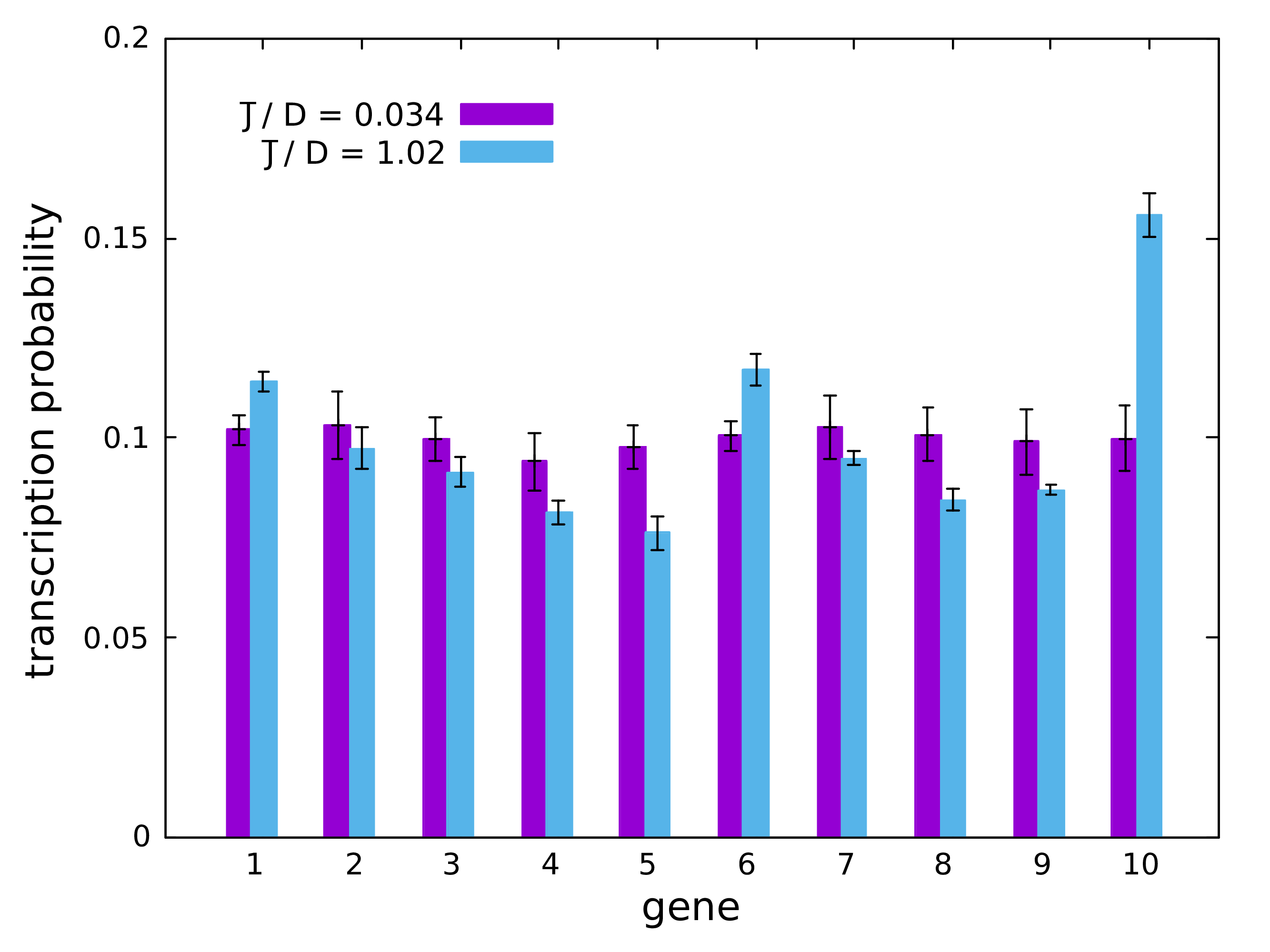}
\caption{\textbf{Transcriptional probability in the tandem 10-gene array.} The histograms show the transcription probability for each gene, for two different values of the flux, $ \bar{J}/D = 0.034 $ and $ \bar{J}/D = 1.02 $, that are the bottom and the top part of the diagram in Fig.~5b in the main text, respectively.}
\label{sfig:figS6}
\end{figure}

In Fig.~\ref{sfig:figS5}a we show a typical time series for the supercoiling at the promoter  of the gene 4 (which is not upregulated), when the burst significance is high ($ \bar{J}/D = 0.68 $, that is in the bursty transcriptional regime). In Fig.~\ref{sfig:figS5}b we show the supercoiling time series in the \textit{wavy} regime: a periodic pattern appears so that the dynamics is no longer bursty. In Figure Fig.~\ref{sfig:figS6} we show the probability of transcription for each gene, in the bursty regime, for two different values of the flux $ \bar{J}/D $. 

\begin{figure}[h]
\centering
\captionsetup[subfloat]{labelfont=bf}
\subfloat[][]{\includegraphics[width=.5\textwidth]{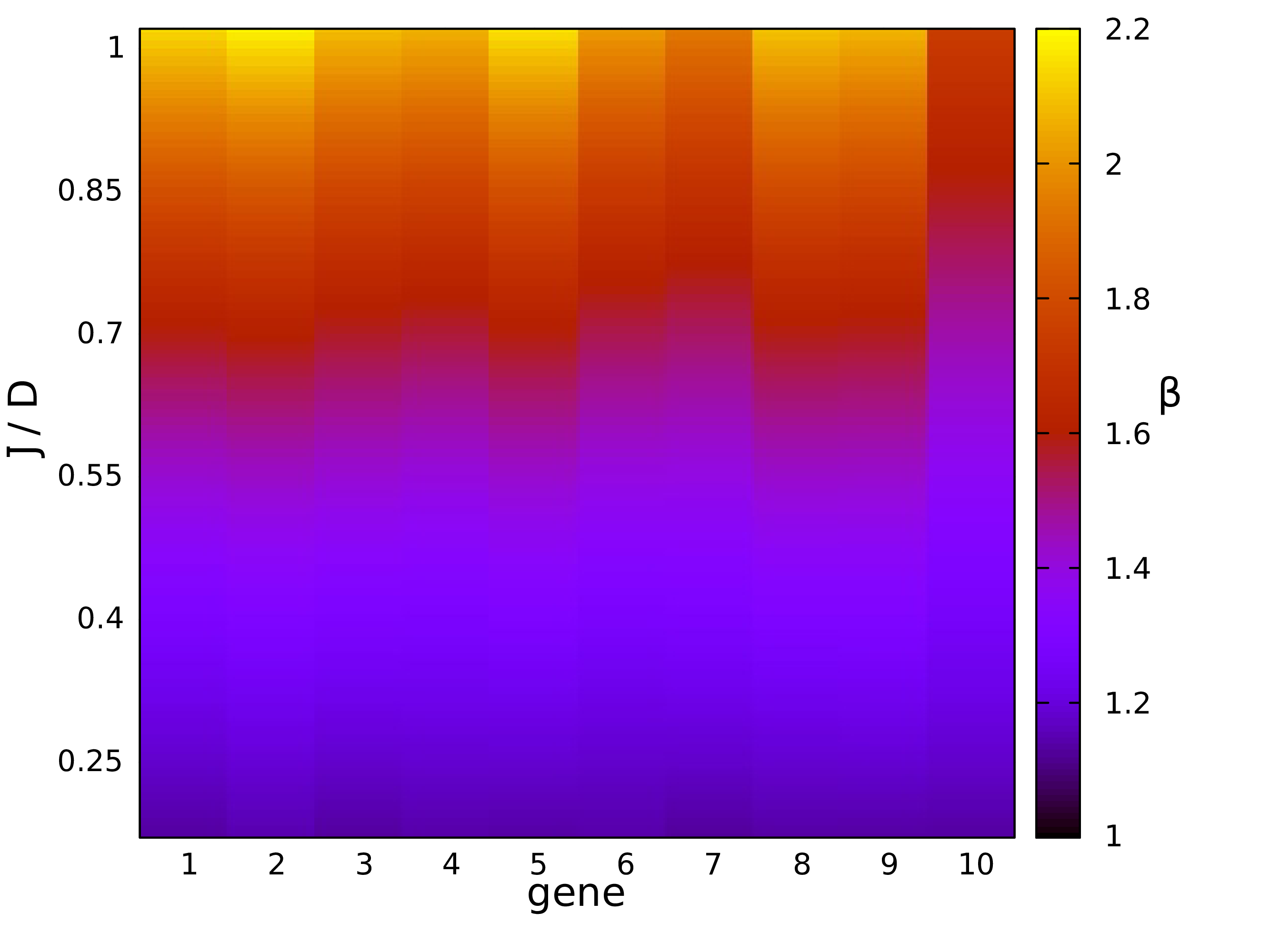}}
\subfloat[][]{\includegraphics[width=.5\textwidth]{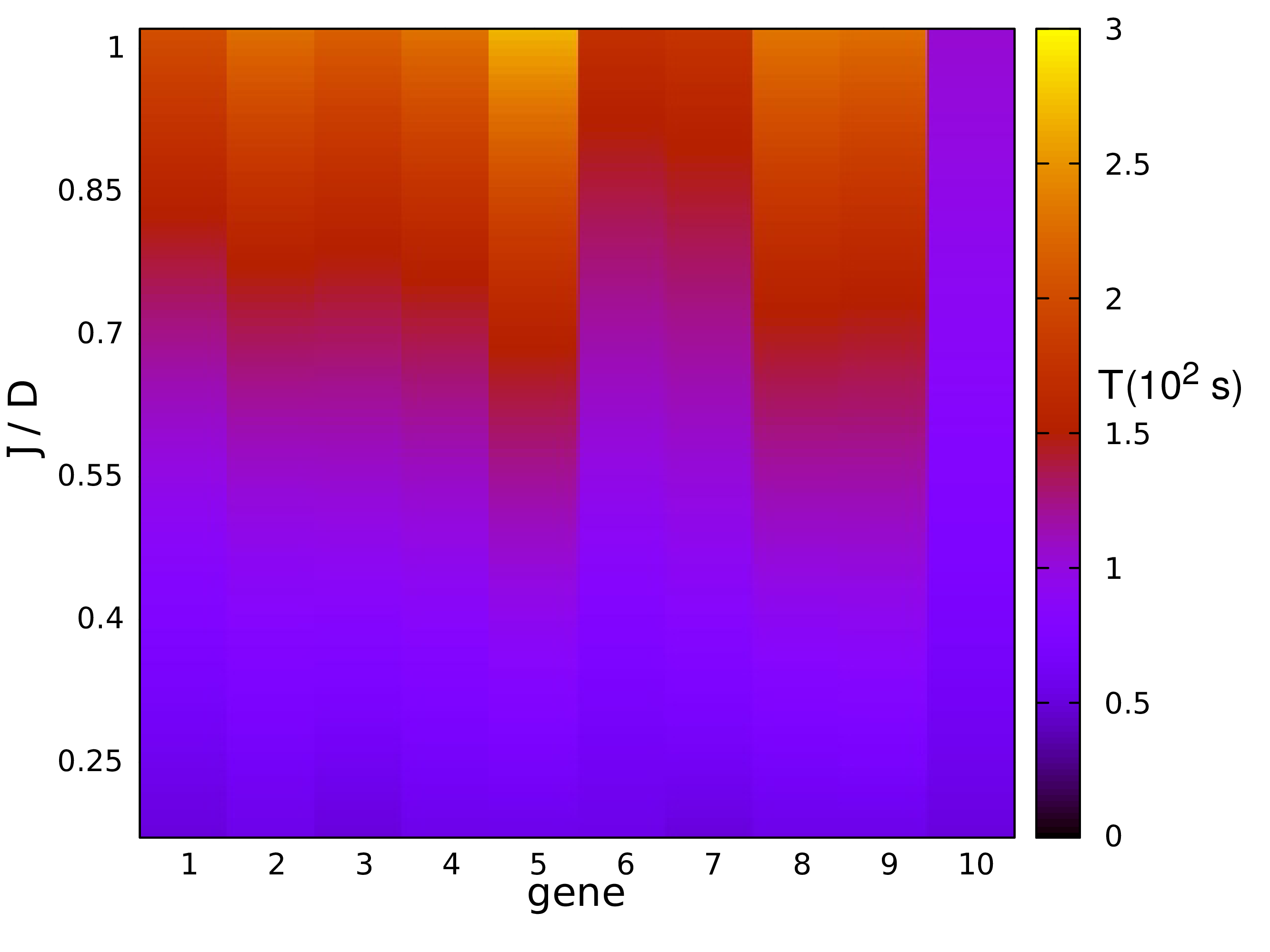}} 
\caption{\textbf{Burst size and burst duration in a 10 genes array}. \textbf{(a)} Burst size $\beta$. \textbf{(b)} Burst duration $T$.}
\label{sfig:figS7}
\end{figure}

In Fig.~\ref{sfig:figS7} we show the mean size of bursts $ \beta $ and the duration of bursts $ T $ for simulations of tandem genes. Even in this case we have a good agreement with the results for a single gene. Indeed we find that in the region of higher burst significance we have $ \beta \gtrsim 2 $ and $ T \sim 4-5 $ min. 

\begin{figure}[b]
\centering
\captionsetup[subfloat]{labelfont=bf}
\subfloat[][]{\includegraphics[width=.5\textwidth]{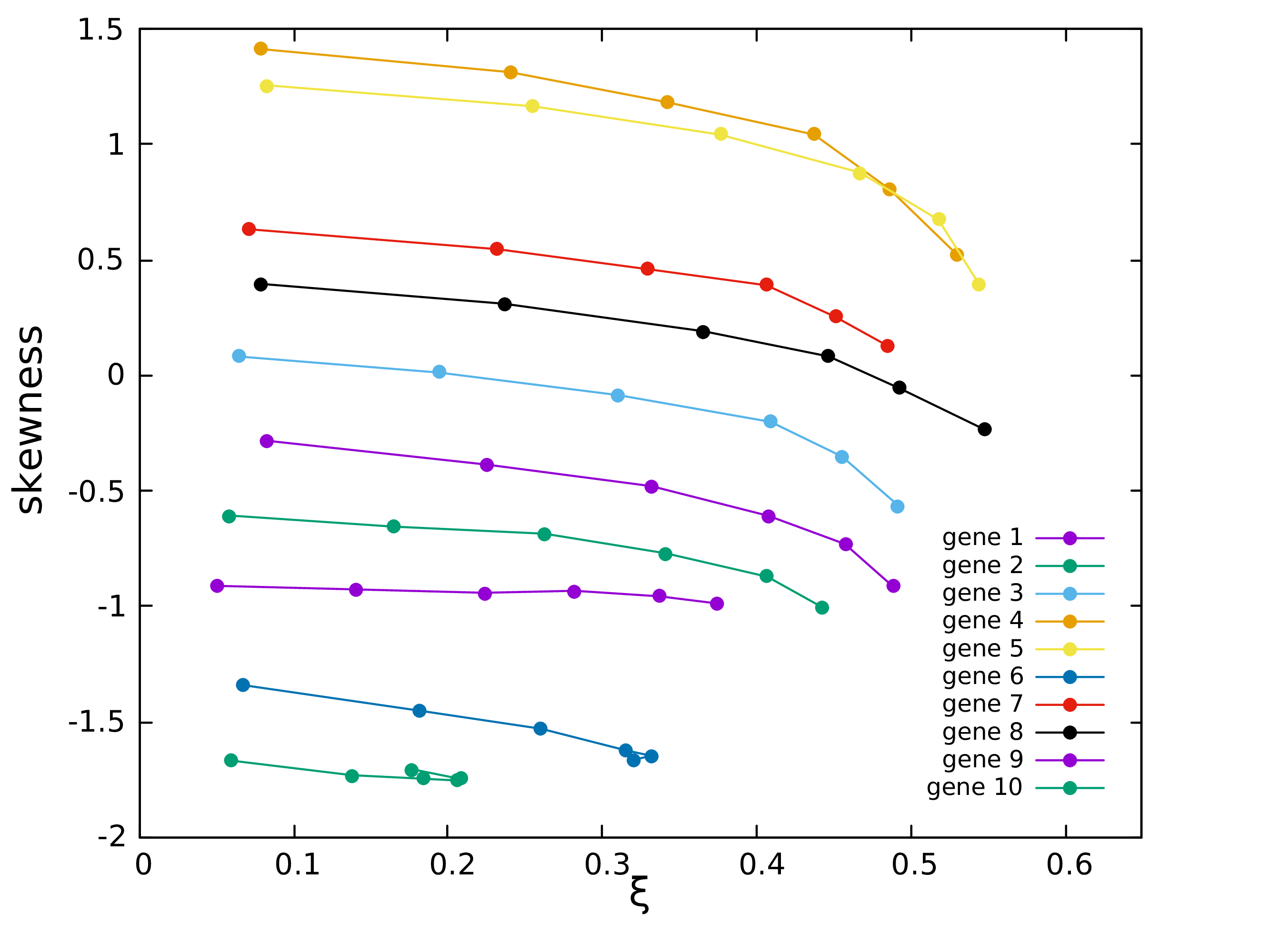}}
\subfloat[][]{\includegraphics[width=.5\textwidth]{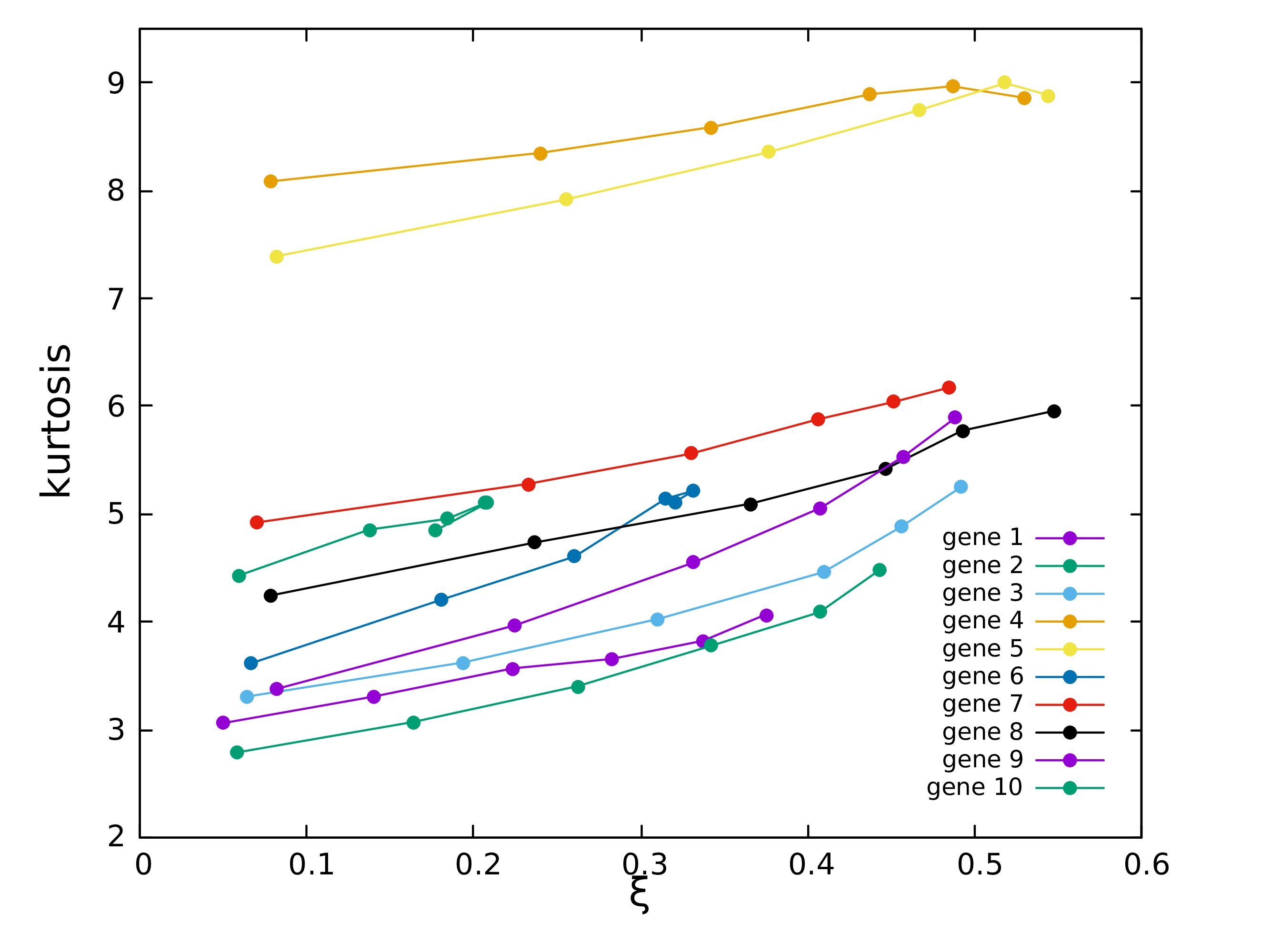}} 
\caption{\textbf{Non-Gaussian parameters for simulations with tandem genes}. \textbf{(a)} Skewness and  \textbf{(b)} kurtosis as a function of $ \xi $. For each gene the slowness and the burst significance are computed for different value of the flux $ \bar{J}/D $. For each gene the skewness decreases as the bursts significance decreases, whereas the kurtosis increases.}
\label{sfig:figS8}
\end{figure}

For completeness, in Fig.~\ref{sfig:figS8} we show the non-Gaussian parameters for the distribution of the supercoiling at the promoter $ \sigma_p $, already computed for a single gene in the main text. The skewness and the kurtosis are not well-correlated to the burst significance, and the values depend strongly on the particular gene considered in a given configuration. However, for each gene individually, the skewness/kurtosis displays a decreasing/increasing trend as a function of $ \xi $. 

\begin{figure}[h]
\centering
\captionsetup[subfloat]{labelfont=bf}
\subfloat[][]{\includegraphics[width=.47\textwidth]{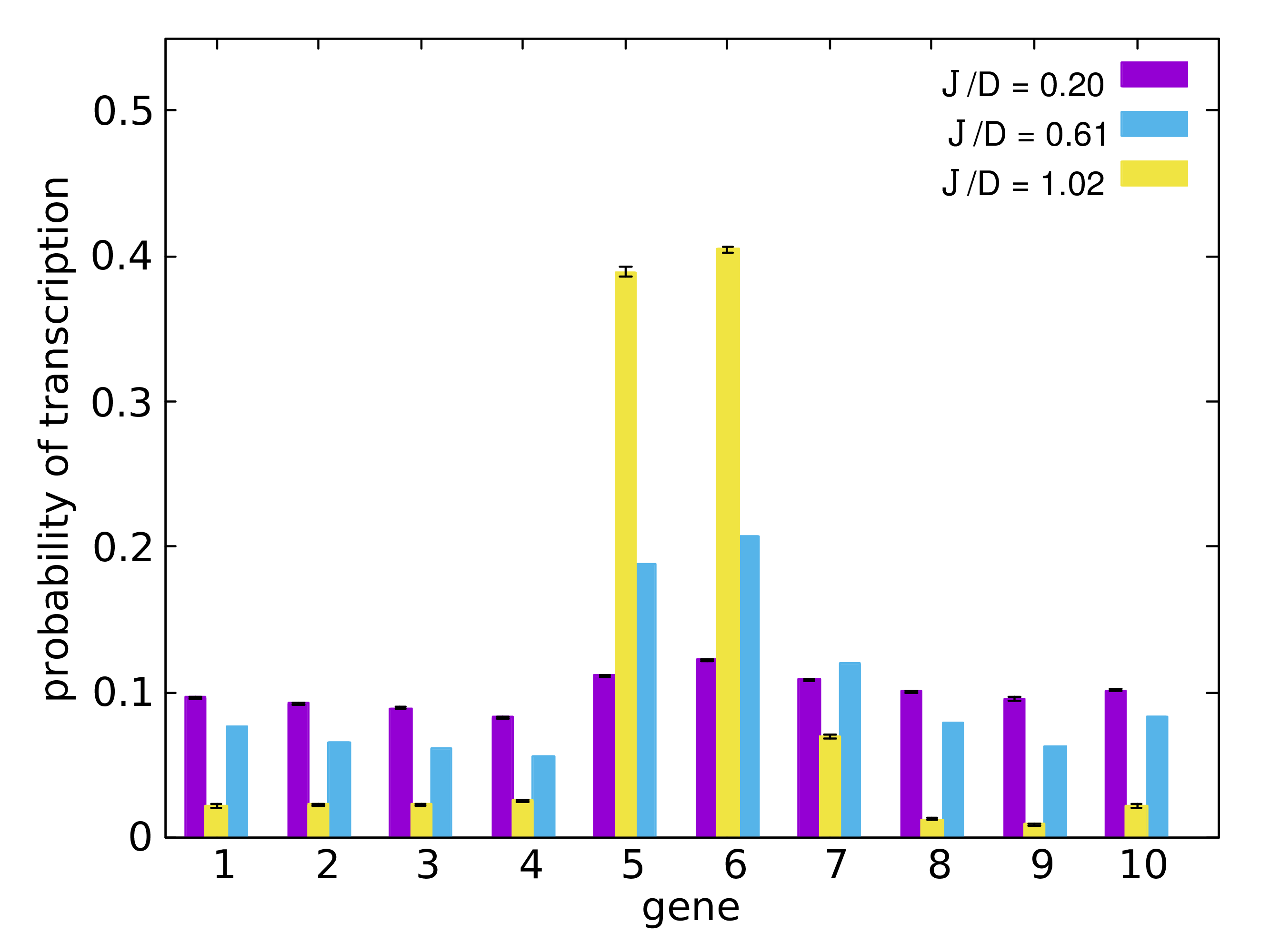}}
\subfloat[][]{\includegraphics[width=.47\textwidth]{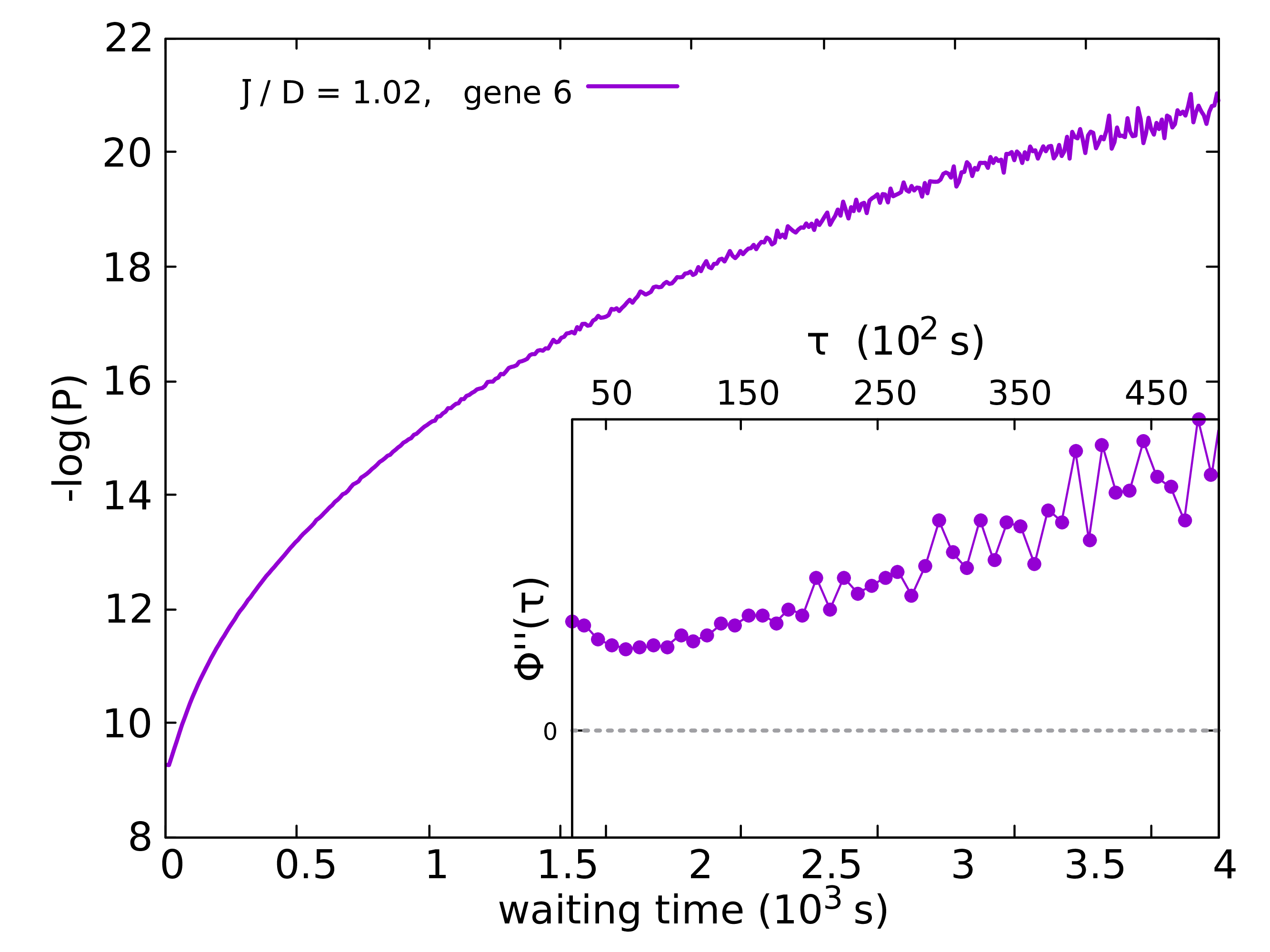}} 
\caption{\textbf{Transcription probability and waiting times distribution for gene 6 in a 10 gene array, with a pair of divergent genes}. \textbf{(a)} Transcription probability for each gene. For small values of the flux ($ \bar{J}/D = 0.2 $, purple boxes) genes are almost equally transcribed. As the flux increases, divergent genes start to dominate the dynamics, and their transcription probability increases, while the others are virtually silenced ($ \bar{J}/D = 1.02 $, yellow boxes). \textbf{(b)} Log-linear plot of the waiting time distribution. The system does not display any bistability. Instead, the system visits several states, each of them described by a particular value of supercoiling at the promoter and a corresponding typical waiting time. Inset: the second derivative of $\Phi(\tau)$ does not display zeros, corresponding with the absence of two separate timescales.}
\label{sfig:figS9}
\end{figure}

In arrays with a pair of divergent genes, the transcription probability for different genes starts to differ as soon as the value of the flux is large enough to give rise to supercoiling mediated interaction (positive feedback loop) between the two divergent genes, Fig.~\ref{sfig:figS9}a. As a consequence, the burst significance $ \xi $ also differs among the genes. Since for high value of the flux ($ \bar{J}/D \sim 1 $) the transcription across all genes is almost totally dominated by the pair of divergent genes, we find that the latter behave like a single upregulated gene. This can be seen by looking at the distribution of waiting times of one of the two genes (Fig.~\ref{sfig:figS9}b), which clearly does not display two separate timescales.

\begin{figure}[b]
\centering
\captionsetup[subfloat]{labelfont=bf}
\subfloat[][]{\includegraphics[width=.48\textwidth]{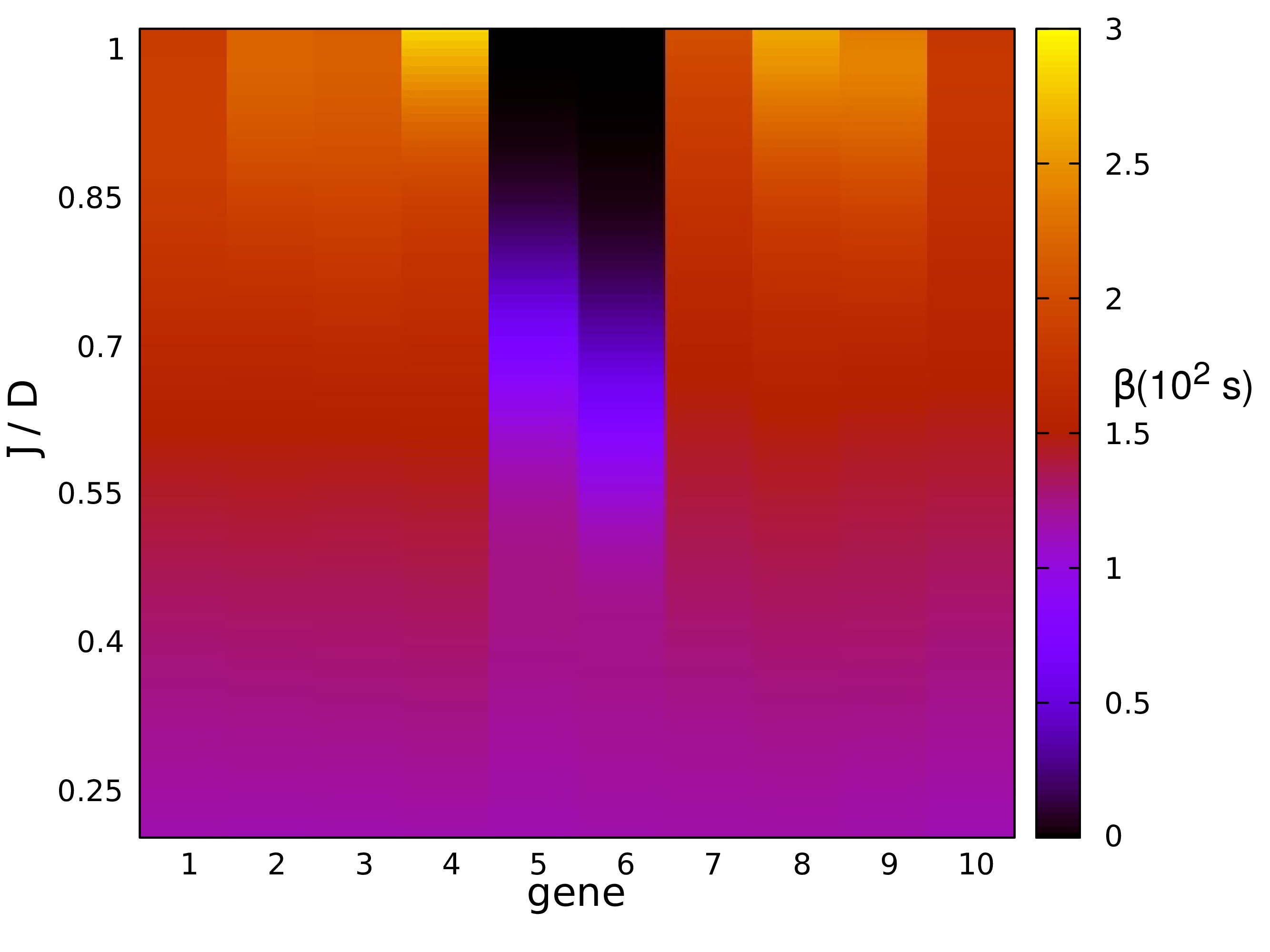}}
\subfloat[][]{\includegraphics[width=.48\textwidth]{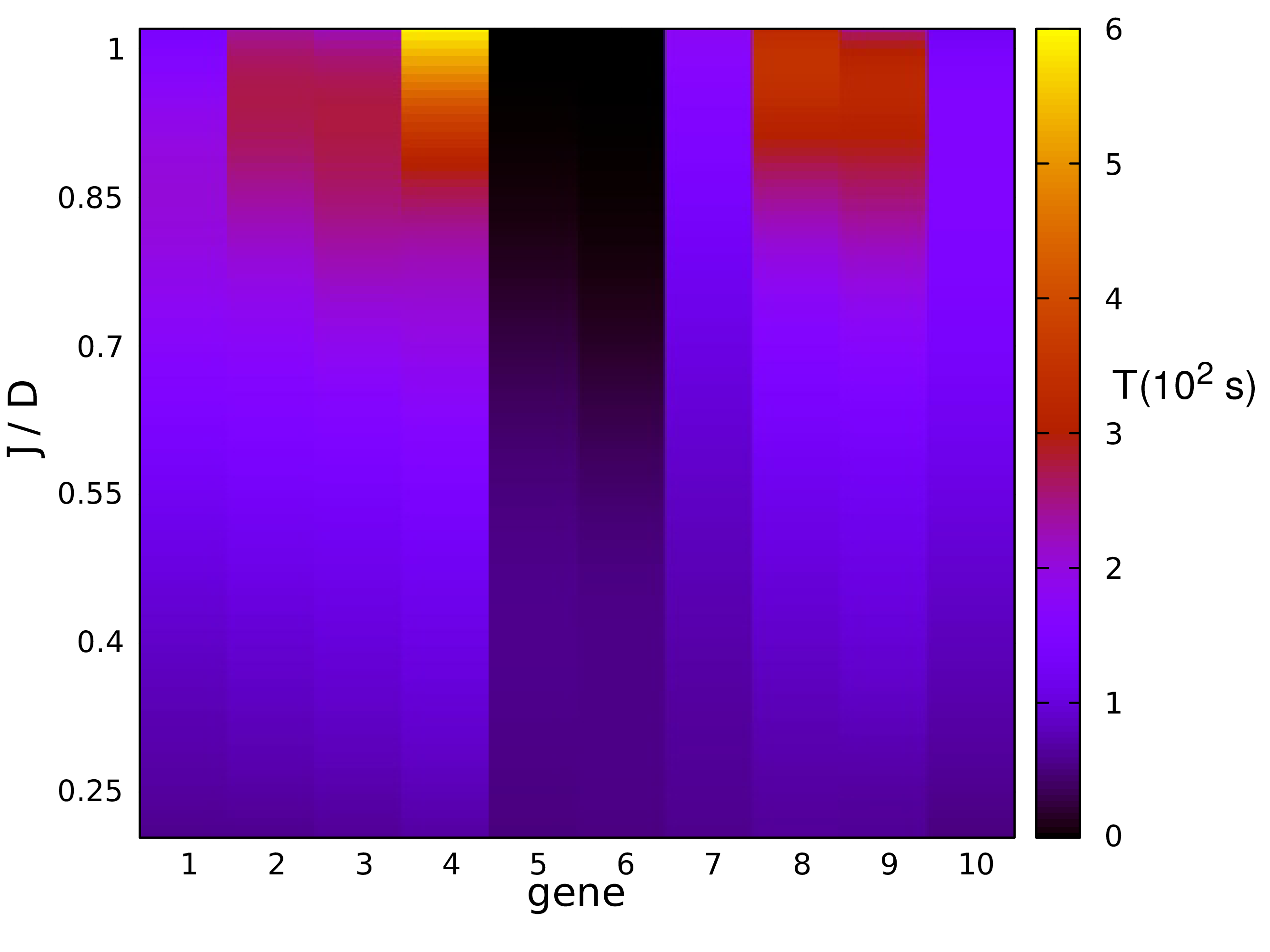}} 
\caption{\textbf{Burst size and burst duration in the presence of divergent genes}. \textbf{(a)} Burst size $\beta$ and \textbf{(b)} Burst duration $T$.}
\label{sfig:figS10}
\end{figure}

\vspace{5mm}
As for the tandem setup, we show the burst parameters $ \beta $ and $ T $ in the presence of a pair of divergent genes (genes $5$ and $6$, see Fig.~\ref{sfig:figS10}). We note that the size of bursts is barely greater than $2$ for high values of $\bar{J}/D$, since bursty genes are strongly down-regulated. 



\begin{thebibliography}{99}
\bibitem{Alberts} B. Alberts, {\it et al.}, {\it Molecular biology of the cell}, Garland Science, New York (2002).
\bibitem{Raj}
A. Raj, and A. van Oudenaarden, {\it Cell}, {\bf 135}, 216 (2008).
\bibitem{Berg} O.~G. Berg, R.~B. Winter, and P.~H. von Hippel,
{\it Biochemistry}, {\bf 20}, 6929 (1981).
\bibitem{Marko}
S. E. Halford and J. F. Marko, { \it Nucleic Acids Res.}, {\bf 32}, 3040 (2004).
\bibitem{Schwanhausser}
B. Schwanh{\"a}usser, D. Busse, N. Li, G. Dittmar, J. Schuchhardt, J. Wolf, W. Chen, and M. Selbach, {\it Nature}, {\bf 473}, 337 (2011).
\bibitem{Golding} I. Golding, J. Paulsson, S. M. Zawilski, and E.~C. Cox, {\it Cell}, {\bf 123}, 1025 (2005).	
\bibitem{Losick}
R. Losick, and C. Desplan, {\it Science}, {\bf 320}, 65 (2008).
\bibitem{Gardiner}
C. W. Gardiner, { \it Handbook of Stochastic Methods: For Physics, Chemistry, and the Natural Science}, Springer, (2004).
\bibitem{Sneppen}
N. Mitarai, I.~B. Dodd, N.~T. Crooks and K. Sneppen, {\it PLoS Comp. Biol.}, {\bf 4}, e1000109 (2008).
\bibitem{Suter} 
D.~M. Suter {\it et al.}, {\it Science}, {\bf 332}, 1198817 (2011).
\bibitem{Harper}
C.~V. Harper {\it et al.}, {\it PLOS Biology}, {\bf 9}, 1000607 (2011).
\bibitem{Levine2018}
S.~A. Sevier, H. Levine, Nucleic Acid Res., gky382 (2018).
\bibitem{Corrigan}
A. M. Corrigan, E. Tunnacliffe, D. Cannon, and J. R. Chubb, { \it eLIFE}, 13051 (2016).
\bibitem{Chong}
S. Chong, C. Chen, H. Ge, and X. S. Xie, {\it Cell}, {\bf 158}, 314 (2014).
\bibitem{Maxwell}
A.~D. Bates, and A. Maxwell, {\it DNA topology}, (Oxford University Press, New York, 2005).	
\bibitem{Brackley}
C.~A. Brackley, J. Johnson, A. Bentivoglio, S. Corless, N. Gilbert, G. Gonnella, and D. Marenduzzo, {\it Phys. Rev. Lett.}, {\bf 117}, 018101 (2016).
\bibitem{Naughton}
C. Naughton, N. Avlonitis, S. Corless, J.~G. Prendergast, I.~K. Mati, P.~P. Eijk, S.~L. Cockroft, M. Bradley, B. Ylstra, and N. Gilbert,
{\it Nat. Struct. Mol. Biol.}, {\bf 20}, 387 (2013).
\bibitem{Kouzine}
F. Kouzine, A. Gupta, L. Baranello, D. Wojtowicz, K. Ben-Aissa, J. Liu, T.~M. Przytycka, and D. Levens, {\it Nat. Struct. Mol. Biol.}, {\bf 20}, 396 (2013).
\bibitem{Chaikin1995} 
P.~M. Chaikin and T.~C. Lubensky, 
{\it Principles of Condensed Matter Physics}, Cambridge University Press
(Cambridge) (1995).
\bibitem{vonLoenhout2012}
M.~T.~J. van Loenhout, M.~V. de Grunt, and C. Dekker, {\it Science}, {\bf 338}, 94 (2012).
\bibitem{Moulin}
L. Moulin, A. Rahmouni, A. Rachid and F. Boccard, {\it Mol. Microbiol.}, {\bf 55}, 601 (2005).
\bibitem{Liu}
L.~F. Liu and J.~C. Wang, {\it Proc. Natl. Acad. Sci.}, {\bf 84}, 7024 (1987).
\bibitem{Marko2007}
  J.~F. Marko, {\it Phys. Rev. E}, {\bf 76}, 021926 (2007).
\bibitem{Liang99}
S.-T. Liang, M. Bipatnath, Y.~C. Xu, S.~L. Chen, P. Dennis, M. Ehrenberg, H. Bremer, {\it J. Mol. Biol.} {\bf 292}, 19 (1999).
\bibitem{yeasttranscription} V. Pelechano, S. Chavez, J.~E. Perez-Ortin,
{\it Plos ONE} {\bf 5}, e15442 (2010).
\bibitem{humantranscription} 
D.~A. Jackson, A. Pombo, F. Iborra, {\it FASEB J.} {\bf 14}, 242 (2000).
\bibitem{Bremer1996}
H.~D. Bremer and P.~P. Dennis, {\it Escherichia coli and Salmonella: cellular and molecular biology}, {\bf 2}, 1553 (1996).
\bibitem{Ishihama2008} Y. Ishihama {\it et al.}, {\it BMC Genomics}, {\bf 9}, 102 (2008).
\bibitem{Terekhova2012} K. Terekhova, K.~H. Gunn, J.~F. Marko and A. Mondragon, {\it Nucl. Acids Res.}, {\bf 40}, 10432-10440 (2012).
\bibitem{Weintraub1986}
H. Weintraub P.~F. Cheng, and K. Conrad, {\it Cell}, {\bf 46}, 115 (1986).
\bibitem{Rhee1999}
K.~Y. Rhee, M. Opel, E. Ito, S. Hung, and G.~W. Hatfield, {\it Proc. Natl. Acad. Sci.}, {\bf 96}, 14294 (1999).
\bibitem{Dobrzynsky}
M. Dobrzy\'nsky, and F.~J. Bruggerman, {\it Proc. Natl. Acad. Sci.}, {\bf  106}, 2583 (2009).
\bibitem{Kumar}
N. Kumar, A. Singh, and R. V. Kulkami, {\it PLoS Comput. Biol.}, {\bf 11}, 1004292 (2015).
\bibitem{Cagnetta}
F. Cagnetta, F. Corberi, G. Gonnella, and A. Suma, {\it Phys. Rev. Lett.}, {\bf 119}, 158002 (2017).
\bibitem{Janas}
M. Janas, A. Kamenev, and B. Meerson {\it Phys. Rev. E}, {\bf 94}, 032133 (2016).
\bibitem{Fukaya}
T. Fukaya, B. Lim, and M. Levine, {\it Cell}, {\bf 166}, 358 (2016).
\bibitem{Dean}
F.~B. Dean, and N.~R. Cozzarelli, {\it J. Biol. Chem.}, {\bf 260}, 4984 (1984).	
\bibitem{Karn}
M. K\ae rn, T.~C. Elston, W.~J. Blake and, J.~J. Collins, {\it Nat. Rev. Genet.}, {\bf 6}, 451 (2005).
\bibitem{McAdams}
H.~H. McAdams, and A. Arkin, {\it Proc. Natl. Acad. Sci.}, {\bf 94}, 814 (1997).
\bibitem{Spudich}
J.~L. Spudich, and D.~E. Koshland Jr, { \it Nature}, {\bf 262}, 467 (1976).
\bibitem{Becskei} 
A. Becskei, and L. Serrano, { \it Nature}, {\bf 405}, 590 (2000).
\bibitem{Thattai}
M. Thattai and A., van Oudenaarden,  { \it Proc. Natl. Acad. Sci.}, {\bf 98}, 8614 (2001).
\bibitem{Ingram}
P.~J. Ingram, M.~P.~H. Stump, and J. Stark, {\it PLoS Comput. Biol.}, {\bf 730}, 1000192 (2008).
\bibitem{Champoux}
J.~J.Champoux, {\it Annu. Rev. Biochem.}, {\bf 70}, 369 (2001).
\bibitem{Dorman}
C.~J. Dorman, and M.~J. Dorman,  { \it Biophys. Rev.}, {\bf 8}, 209 (2016).
\bibitem{Kamien}
R.~D. Kamien, {Eur. Phys. J. B}, {\bf 1} (1998)

\end{thebibliography}

\clearpage

\begin{thebibliography}{99}
\bibitem{Brackley}
C.~A. Brackley, J. Johnson, A. Bentivoglio, S. Corless, N. Gilbert, G. Gonnella, and D. Marenduzzo, {\it Phys. Rev. Lett.}, {\bf 117}, 018101 (2016).
\bibitem{Dobrzynsky}
M. Dobrzy\'nsky, and F.~J. Bruggerman, {\it Proc. Natl. Acad. Sci.}, {\bf  106}, 2583 (2009).
\bibitem{Kumar}
N. Kumar, A. Singh, and R.~V. Kulkami, {\it PLoS Comput. Biol.}, {\bf 11}, 1004292 (2015).
\bibitem{Golding} I. Golding, J. Paulsson, S.~M. Zawilski, and E.~C. Cox, {\it Cell}, {\bf 123}, 1025 (2005).
\bibitem{Chong}
S. Chong, C. Chen, H. Ge, and X.~S. Xie, {\it Cell}, {\bf 158}, 314 (2014).
\end{thebibliography}
\end{document}